\documentclass[prd,aps,superscriptaddress,floatfix,nofootinbib,eqsecnum,twocolumn]{revtex4-2}

\pdfoutput=1

\usepackage{amsfonts}
\usepackage{amsmath}
\usepackage{amssymb}
\usepackage{bm}
\usepackage{mdframed}
\usepackage{xcolor}
\usepackage{dcolumn}
\usepackage{graphicx}
\usepackage[latin1]{inputenc}
\usepackage{latexsym}
\usepackage{rotating}
\usepackage{hyperref}
\usepackage{subfigure}
\usepackage{color}
\usepackage{changes}
\usepackage{verbatim}
\usepackage{comment}
\usepackage{empheq}
\usepackage{csquotes}
\usepackage{physics}
\usepackage{float}
\usepackage{amsmath,latexsym}
\usepackage{booktabs}  
\usepackage{soul}
\begin{document}
 \newcommand{\hsp}{\hspace{0.1mm}}
 \newcommand{\bq}{\begin{equation}}
 \newcommand{\eq}{\end{equation}}
 \newcommand{\bqn}{\begin{eqnarray}}
 \newcommand{\eqn}{\end{eqnarray}}
 \newcommand{\nb}{\nonumber}
 \newcommand{\lb}{\label}
 \newcommand{\pp}{\partial}


\title{
GUP deformed background dynamics of phantom field
}

\author{Gaurav Bhandari}\email{bhandarigaurav1408@gmail.com}\affiliation{Department of Physics, Lovely Professional University, Phagwara, Punjab, 144411, India}

\author{S. D. Pathak}\email{shankar.23439@lpu.co.in}\affiliation{Department of Physics, Lovely Professional University, Phagwara, Punjab, 144411, India}

\author{Manabendra Sharma}\email{sharma.man@mahidol.ac.th}\affiliation{Centre for Theoretical Physics and Natural Philosophy, Nakhonsawan Studiorum for Advanced Studies, Mahidol University, Nakhonsawan, 60130, Thailand}

\author{Anzhong Wang}\email{Anzhong$_$Wang@baylor.edu}\affiliation{GCAP-CASPER, Physics Department, Baylor University, Waco, TX 76798-7316, USA}

\begin{abstract}
Quantum gravity has been baffling the theoretical physicist for decades now, both for its mathematical obscurity and phenomenological testing. Nevertheless, the new era of precision cosmology presents a promising avenue to test the effects of quantum gravity.
In this study, we consider a bottom-up approach. Without resorting to any candidate quantum gravity, we invoke a generalized uncertainty principle (GUP) directly into the cosmological Hamiltonian for a universe sourced by a phantom scalar field with  potential to study  the evolution of the universe in a very early epoch. This is followed by a systematic analysis of the dynamics, both qualitatively and quantitatively. Our qualitative analysis shows that the introduction of GUP significantly alters the existence of fixed points for the potential considered in this paper. In addition, we confirm the existence of an inflationary phase and analyze the behavior of relevant cosmological parameters with respect to the strength of the GUP distortion.





\end{abstract}

\maketitle



\section{
Introduction
}
\label{Introduction}
The advent of Einstein's general theory of relativity 
has given birth to many fields of research. Since gravity is the only dominant force at large distances, general relativity (GR) provides a viable mathematical framework to construct models of cosmology. Over a period of hundred years, GR has seen profound successes. A few classical examples include the explanation of the perihelion precession of Mercury\cite{rana1987investigation}, the deflection of light rays when passing close by massive bodies \cite{genov2009mimicking}, and the gravitational redshift of light \cite{wojtak2011gravitational}.

In particular, to cosmology, in the year 1929 the discovery of the Hubble's expansion law laid the foundation of modern cosmology.
This observational evidence of uniform and isotropic expansion of the universe as incorporated by the Friedmann-Lemaitre-Robertson-Walker (FLRW) universe gives rise to the standard model of cosmology (SMC). The FLRW metric is a maximally symmetric geometry of spacetime that supports the Copernican principle. One of the remarkable successes of SMC is the prediction of cosmic microwave background radiation (CMB).
%
Although successful, however, the SMC has been confronted with some serious drawbacks. An example is the so-called horizon problem, that is, the causal explanation for two otherwise spatially disconnected regions of space is lacking within the scope of the SMC. Others  are   flatness and entropy problem \cite{albrecht1982cosmology,guth1981inflationary,liddle1992cobe,kinney2009tasi,riotto2002inflation}.

The inflationary paradigm proposed by A. Guth (1981) rescues the situation by providing a mechanism to solve the puzzles of the SMC with the help of a nearly exponential expansion of the universe at a very early stage. A scalar field with a proper potential serves as a good candidate for the inflationary scenario.

The inflationary epoch not only rescues SMC but also predicts the formation of the large-scale structure of the universe. Although the universe looks almost homogeneous and isotropic at large scale \cite{A.A.Starobinsky}, the tiny fluctuation of the order of $10^{-5}$ has been observed in CMB. This tininess of the scale allows us to employ perturbation theory, wherein the zeroth order, the background of the spacetime,   is still FLRW and any inhomogeneity is given by the leading order correction. The physical reason for the perturbation of spacetime is the quantum fluctuation of matter content, which is the inflaton in the current situation. Of course, any perturbation of the matter field would induce a perturbation in the gravitational field, resulting in the clumping of energy and matter density, leading to the formation of the large-scale structure we see today. In the process, inflation expands the tiny causally connected quantum fluctuations into the super-Hubble modes, which re-enter the Hubble radius at  later epochs, giving us a causal mechanism for the large-scale structure \cite{guth1982fluctuations,linde1982new,bardeen1983spontaneous}.

In the context of inflation, the homogeneous and isotropic universe is still treated classically while quantizing only the first-order corrections in the linearized theory of gravity. However, as the scale approaching to the Planck regime \cite{ashtekar2005quantum,ashtekar2009loop},  one would expect the quantum nature of the background to play a significant role.
This incomplete picture of the theory of cosmology, at present, is due to the continued lack of a consistent candidate for quantum gravity. In fact, this is one of the most challenging issues in modern physics. The main challenge comes from our current understanding of the nature, based on two mathematically incompatible frameworks:  GR and quantum mechanics (QM)   \cite{c1,s1,s2,h1}.

In the literature, there exist different candidates based on different philosophical approaches to quantize gravity, each with its own advantages and issues. The two major streams of quantum gravity (QG) are string/M theory and loop quantum gravity (LQG). While string/M theory is based on the unification of gravity with three other fundamental forces, LQG is the quantization of the Riemannian geometry of GR  only \cite{ashtekar2021short,rovelli2008loop,rovelli1998strings,rovelli1998strings,Thiemann_2007}. LQG is background independent and non-perturbative. The techniques of LQG, when applied to cosmological spacetime, gives rise to various models of quantum corrected cosmology, also called loop quantum cosmology (LQC) \cite{garay1995quantum}. One of the striking features of LQC is the supplant of initial singularity by quantum bounce owing to the quantization of geometry \cite{Ashtekar:2011,LSW21,LS23,AWW23}. In lieu of its endeavor to empirically grasp the semi-classical physics near the Planck region, LQC is also consistent with observations, and may provide some mechanism to alleviate the anomalies observed currently in cosmology \cite{AGJS20,AKS20}.


Nevertheless, in view of the continued absence of a consistent theory of QG, radically different paths have been adopted. The generalized uncertainty principle (GUP) is one such attempt that can generate quantum corrected dynamics when applied to cosmology to study the very early universe.  In this approach, we consider  the space-time as a probability density associated with basis vectors with additional fluctuations in geometry, giving rise to the extended generalized uncertainty principle (EGUP) \cite{kempf1994quantum,kempf1997quantum}.

The departure point from classical mechanics to the standard quantum mechanics is the Heisenberg's uncertainty principle (HUP), which states the incompatibility of position and momentum operators, reflecting the inherent imprecision of the measurement of one when the other is known precisely.
However, at scales approaching the Planck length, theories of quantum gravity suggest that the geometry of a space-time cannot be measured below the Planck scale. 
Certainly, different physics gives rise to different  minimal lengths. For example,  the minimal length scale   string/M theory   is the string length itself \cite{veneziano1986stringy,witten1996reflections,scardigli1999generalized,gross1988string,amati1989can,yoneya1989interpretation,as01,h56,scardigli2018modified}.
%
This immediately implies that HUP is not applicable at the Planck scale as it puts no limit to precisely measure length provided momentum is undetermined. The inconsistency in HUP indicates the need to modify the existing canonical HUP by incorporating gravitational correction.


The consideration of quantum fluctuation in the space-time geometry leads to GUP, which describes the limitation of measurement of position and momentum. The uncertainties of position and momentum depend on the fluctuation of spacetimes. The greater the uncertainty in the geometry of space, the greater the uncertainty in the position and momentum of the particles \cite{tawfik2015review,lake2021generalised}.
The notion that gravity might influence the uncertainty principle was first proposed by Mead \cite{mead1964possible}. Later, candidate theories of QG such as string/M Theory \cite{s3}, Doubly Special Relativity (DSR) Theory and Black Hole Physics \cite{bosso2021deformed}, introduced modifications to the commutation relations between position and momentum, which are known as the GUP \cite{PhysRevD.85.104029,ADLER_1999}.

In view of the current status of quantum cosmology, the GUP-modified cosmological dynamics require more attention than before to extract the low-energy regime of QG. In particular, in this article, we consider a toy model consisting of phantom scalar field, first with a positive cosmological constant and then with an arbitrary potential for the given GUP.
Phantom inflation leads to a cosmological scenario of the Big Rip, where the universe undergoes a catastrophic expansion that leads to tearing apart all the bound structures, including planets, galaxies, stars, and even fundamental particles.  However, investigating these consequences helps in understanding the possible fate of our universe. Lately, there has been significant attention to phantom cosmology, see, for example, Refs. \cite{q1,q2,q3,q4,sahni2000case,chimento2004big,piao2003nearly,gonzalez2003wormholes}.

In this paper, in Sec.\ref{GUP} we start with reviewing the  formulation of GUP-corrected Hamiltonian, starting with the Einstein-Hilbert action  together with a minimally coupled phantom scalar field and a positive cosmological constant. We obtain the GUP-corrected Friedmann, Raychaudhuri   and Klien-Gordon equations. In  Sec.\ref{potential}, these are extended to include tan arbitrary potential of the scalar field. The techniques of dynamical system analysis have been employed to extract qualitative information about the system in Sec.\ref{DSA}. We limit ourselves to quadratic and exponential potentials. In Sec.\ref{quadratic}, we study inflationary dynamics by calculating the Equation of State (EoS) and the slow climb parameters and plot them out explicitly for quadratic and exponential potentials in Sec.\ref{expo.}.

\section{ GUP-modified background dynamics}

SMC is based on the ``Copernican Principle", which says that the universe is homogeneous and isotropic on a large scale. This is encoded in the maximally symmetric flat FLRW  universe
\begin{equation}\label{metric}
ds^2= -N^2(t)dt^2 + a^2(t)\left[dr^2 +  r^2 d\Omega^2\right].
\end{equation}
Since   GR is a field theory, its dynamics can be obtained from the Euler-Lagrange equation by varying the metric and the matter field of the Einstein-Hilbert action:
\begin{equation}\label{Action}
S_{EH}= \frac{1}{2\kappa}\int d^4x \sqrt{-g} R + \mathcal{L}_m,
\end{equation}
to give the Einstein's equations $G_{\mu \nu}= \kappa T_{\mu \nu}$,
Where $G_{\mu \nu}$ is the Einstein Tensor, $T_{\mu \nu}$  the energy momentum tensor, and $\kappa\equiv {8\pi G}/c^4$ is set equal to one for the rest of the paper owing to the usage of the natural units.

The discovery of late time acceleration of the universe has led to   two major approaches to address the issue of the acceleration. One is based on modifying the gravity sector and the other is on the matter sector  \cite{ratra1988cosmological,peebles2003cosmological,chimento1996scalar,copeland2006dynamics}.
%
In this paper, we focus on the second approach, that is, to modify the matter content to address the early epoch of the universe. To this effect, we adopt a phantom scalar field. This has been extensively studied in the context of the late-time era of evolution. The fact that the phantom field produces a phase of accelerated expansion of the universe makes it interesting to investigate its implication in inflationary dynamics as well.

One of our prime focuses is to explore the tail-end dynamics of the universe where QG effects is still important but not necessarily dominant. In the domain of LQC, this has been reported as \textit{transition phase} from quantum to the classical universe in pre-inflationary dynamics \cite{zhu2017pre,li2018qualitative,sharma2019background}.
However,  this paper will take a radically different approach by directly invoking a GUP in the cosmological Hamiltonian. This is an effective way of modeling the quantum corrected background evolution of the universe.
In this section and in what follows, we review the construction of the GUP-deformed background equation of motion for a phantom scalar field.



\subsection{
Phantom scalar field with cosmological constant
}
\label{GUP}

\subsubsection{Classical Dynamics}
In this section, we consider the Einstein--Hilbert action with a minimally coupled phantom scalar field and a positive cosmological constant,
{
\begin{equation}
S_{\text{EH}} = \int \sqrt{-g} \left[ \frac{1}{2\kappa}\left(R-2\Lambda \right) +\frac{1}{2}g^{\mu \nu}\partial_{\mu}\phi\partial_{\nu}\phi - V(\phi)\right]d^4x,
\end{equation}
on the background of a maximally symmetric spacetime described by Eq.(\ref{metric}), where $V(\phi)$ is the potential of the scalar field $\phi$.
Given the flat FLRW background, our action takes the following form
\begin{equation}
S_{\text{EH}}= V_0 \int dt \left[ -\frac{3a\dot{a}^2}{N}-a^3\left(\frac{\dot{\phi}^2}{2N}+N\left(\Lambda + V\right) \right)\right], \label{cosmo}
\end{equation}
where $V_0$ is the  volume of a fiducial cells, introduced to facilitate our calculations in a non-compact flat FLRW spacetime.
Later, we can take the limit $\lim V_0 \rightarrow \infty$, as the final results will be independent of its values. Therefore, for the sake of simplicity,  
it can be set to 1 without loss of generality. Recall that in this paper we choose the natural units so that $\kappa=1$. }

Thus,   from Eq.(\ref{cosmo}) we can see that the the Lagrangian density is given by
\begin{equation}\label{FreePhantomLagrangian}
\mathcal{L}=-\frac{3a\dot a^2}{N}-a^3\left(\frac{\dot\phi^2}{2N}+N\left(\Lambda + V\right)\right),
\end{equation}
from which we can see that $\mathcal{L}$  does not depend on $\dot{N}(t)$. Hence there is no dynamics in the lapse function $N(t)$, as now we have $P_{N} \equiv \frac{\partial \mathcal{L}}{\partial \dot{N}}=0$. Therefore, the dynamics of the system are completely contained in the equations of motion for $(a, P_a, \phi, P_{\phi})$ governed by the Hamiltonian
\begin{equation}
\mathcal{H}=-N\left(\frac{P_a^2}{12a}+\frac{P_\phi^2}{2a^3}-a^3\left(\Lambda + V\right)\right),\label{Hamiltonian}
\end{equation}
which is obtained from Eq.(\ref{FreePhantomLagrangian}) through the Legendre transformation, where $P_a\equiv \frac{\delta \mathcal{L}}{\delta \dot{a}}$ and $P_{\phi}=\frac{\delta \mathcal{L}}{\delta \dot{\phi}}$ are the conjugate momentum to $a$ and $\phi$, respectively,
with the symplectic structure,
\begin{eqnarray}\label{Symplectic}
\{a,P_a\}   = 1,\quad
\{\phi, P_{\phi}\} = 1.
\end{eqnarray}

Then, the corresponding Friedmann, Raychaudhuri and Klein-Gordon equations (see appendix B) are given respectively by 
\begin{eqnarray}
\lb{Feq}
&& 3H^2 = -\frac{\dot \phi^2}{2}+\Lambda + V(\phi), \\
\lb{Req}
&& 2\frac{\Ddot{a}}{a^2}+ \left(\frac{\Dot{a}}{a}\right)^2= \frac{\Dot{\phi}^2}{2} +\Lambda +V(\phi),\\
\label{Vklein}
&& \Ddot{\phi}+3\Dot{\phi}\left(\frac{\Dot{a}}{a}\right)-\frac{dV(\phi)}{d\phi}=0.
\end{eqnarray}.




\subsubsection{GUP deformed dynamics} \label{GUP deformed}

In this subsection, we review the inclusion of higher-order correction of the uncertainty principle in the cosmological Hamiltonian without the potential $V(\phi)$. To achieve our goal, we first perform a canonical transformation of the phase space 
 in $x$ and $y$ variables such that our Hamiltonian gets simplified and making it easier to incorporate the effects of GUP as follows:
\begin{equation}\label{CanonicalTransformation}
x= \frac{a^{3/2}}{\mu}\sin(\mu \phi),  \hspace{0.5cm} y= \frac{a^{3/2}}{\mu}\cos(\mu \phi),
\end{equation}
while preserving the dynamics. 
From Eq.(\ref{CanonicalTransformation}), with  $\mu=\sqrt{3/8}$ we obtain the following
{
\bqn
\label{FEq}
&& \dot x ^2 \sin^2(\mu\phi)+\dot y^2 \cos^2(\mu\phi)+2\dot x \dot y \sin(\mu\phi)\cos(\mu\phi) =6a\dot a^2, \nb\\
&&\dot x ^2 \cos^2(\mu\phi)+ \dot y^2 \sin^2(\mu\phi)-2\dot x \dot y \cos(\mu\phi)\sin(\mu\phi)= a^3\dot \phi^2,\nb\\
\eqn
and 
\bq
\label{Circle}
x^2+y^2 = \frac{a^3}{\mu^2} \sin^2(\mu\phi)+ \frac{a^3}{\mu^2} \cos^2(\mu\phi) = \frac{8a^3}{3}.
\eq}
Now, from Eq.(\ref{Circle}) we observe that the physical volume of the universe under study can be elegantly expressed as the radius of the circle, with $(0,0)$ as the center in the plane containing the configuration variable $(x,y).$ Since the Friedmann equation is nothing but the fractional rate of change of volume, intuitively, one can speculate that the knowledge of the dynamics of the pair $(x,y)$ suffices to predict the evolution of the universe.

Having set the stage in terms of the Cartesian pair (x,y), we now return to the question of dynamics.  
We obtain the final form of the Lagrangian in terms of configuration coordinates $(x,y,\dot{x},\dot{y})$ to be
\begin{equation}
\mathcal{L}= -\left[\frac{\dot x^2 + \dot y^2}{2N} + \frac{3}{8}(x^2+y^2)\Lambda N \right].
\end{equation}
A quick look at the form of the Lagrangian suggests that dynamics is symmetric w.r.t the origin of the circle in the plane of $(x,y)$.

Given this, it is straightforward to obtain the canonically transformed Hamiltonian using the Legendre transformation.
Thus, the final form of the Hamiltonian takes the form
\begin{equation}
\mathcal{H}_0= N\left[\frac{P_x^2}{2}+\frac{P_y^2}{2}+\frac{\omega^2}{2}(x^2+y^2)\right]\label{HClassicalFree},
\end{equation}
where  we set  $\omega^2=-\frac{3}{4}\Lambda$, {\em which, formally, represents the frequency of two independent harmonic oscillators}. The utility of the canonical transformation of Eq.(\ref{CanonicalTransformation}) is clear from the elegant expression of Eq.(\ref{HClassicalFree}), and the dynamics of a universe with a phantom scalar and a positive cosmological constant can be expressed as a system of two decoupled simple harmonic oscillators in the new phase space representation.

However, Eq. (\ref{HClassicalFree}) is still classical, though expressed in a different form. The minimal uncertainty in position at the Planck scale is a feature present in many QG theories \cite{bos1}. On the other hand, the HUP, one of the pillars of QM, allows for an arbitrary value of uncertainty in position as long as the product of the position and momentum uncertainties is larger than $\frac{\hbar}{2}$ \cite{sc1,k1,ben1}. With the existing HUP, we have an arbitrary choice of uncertainty, and hence precision  in position, which motivates modifying the Heisenberg relation. One of the most notable and heuristic approaches is the   GUP, first introduced by Kempf, Mangano, and Mann (KMM) in 1995, where the uncertainty relation in one dimension is obtained via the deformed commutator bracket
\begin{equation} \label{KMM}
[x, p] = i \hbar (1+\alpha^2 p^2).
\end{equation}
Later, a more general form of GUP was proposed by Ali, Das, and Vagenas \cite{al1} in 2011, which incorporates linear and quadratic dependence on momentum in different QG theories. The presence of the quadratic term is dictated by String/M Theory\cite{am2,gr1} and black hole physics \cite{mag1,scar,sc9,sc10}, while the linear momentum-dependent term is motivated by Doubly Special Relativity (DSR) \cite{am1}. The GUP deformed commutator bracket is expressed as
\begin{equation}\label{das}
[q_i,p_j]=i\hbar\Bigg\{\delta_{ij}- \alpha \left(p\delta_{ij}+\frac{p_ip_j}{p}\right) + \alpha^2[ p^2 \delta_{ij}+3p_ip_j]\Bigg\},
 \end{equation}
 where $\alpha= \alpha_0/M_{Pl}c= \alpha_0 l_{Pl}/\hbar$, $M_{Pl}$ = Planck mass, $l_{Pl} \approx 10^{-35} m = \text{Planck length}$, and $M_{Pl}c^2  \approx 10^{19} GeV = \text{Planck energy}$.

 To write down the dynamics due to momentum deformation owing to  GUP, we introduce the semi-classical canonical variables $q_i$ and $P_i$ and GUP in the WDW equation in our cosmological model as followed by \cite{lopez2018phase,ca1}.
The modified commutator relation incorporating the linear and quadratic dependence of momentum \cite{bos1,al1} is
 \begin{equation}\label{used}
    [q_i,P_j]= i \hbar \delta_{ij} (1-2\beta \gamma P_0+4\epsilon \gamma^2 P_0^2).
 \end{equation}
Using this commutator relation, we find that the canonical variable is approximately expressed as
\begin{equation}\label{canonical}
q_i=q_{0i} ,\quad P_i=P_{0i}\left(1-\beta\gamma P_0 +2\gamma^2\frac{\beta^2+2\epsilon}{3}P_0^2\right),
\end{equation}
where $P_0^2=P_{0j}P_{0j} $, $\gamma$ is related to the scales where quantum-gravitational effects became relevant, typically defined by a value proportional to the inverse of the Planck momentum as $\gamma \equiv \frac{\gamma_0}{M_{Pl} c}$ with $\gamma_0 \sim 1$. The parameters  $\beta$ and $\epsilon$ are dimensionless
and highlight the terms originating from linear and quadratic contribution to GUP. For  $\beta =0$, we recover the KMM GUP model.

Now we calculate the GUP distorted Hamiltonian up to the order of $\gamma^2$ \cite{lopez2023generalized}, which is
\begin{widetext}
\begin{equation}
\mathcal{H}= \mathcal{H}_0  -\beta\gamma(P_{0x}^2+p_{0y}^2)^{3/2} +\gamma^2(P_{0x}^2+p_{0y}^2)^2\left(\frac{\beta^2}{6}+\frac{2\epsilon}{3}\right)+ \mathcal{O}(\gamma^3)\label{HgupFree},
\end{equation}
\end{widetext}
where 
\bq
\mathcal{H}_0\equiv\frac{P_{0x}^2}{2}+\frac{P_{0y}^2}{2}+\frac{\omega^2}{2}(x^2+y^2),
\eq
is the unperturbed Hamiltonian before introducing GUP. From now on, the subscript $0$ will be used to denote the unperturbed version of Eq.(\ref{HClassicalFree}). For example, the unperturbed $x,y$ are represented as $(q_{0x},q_{0y})$ while the unperturbed pair $(P_x,P_y)$ as $(P_{0x},P_{0y}).$

In the final step, we re-express the GUP deformed Hamiltonian Eq.(\ref{HgupFree}) in terms of the cosmological phase-space variables. This is achieved by applying the inverse transformation to express Eq(\ref{HgupFree}) in the cosmological variables, namely, the expansion factor, the scalar field, and their corresponding conjugate momenta, where 
\bqn
&& P_{0x} \equiv \dot{x}= \frac{3}{2}\frac{a^{1/2}\dot a}{\mu} \sin (\mu\phi)+(a^{3/2}\dot\phi) \cos(\mu\phi),\label{Px}\\
&& P_{0y}= \dot{y}= \frac{3}{2}\frac{a^{1/2}\dot a}{\mu} \cos (\mu\phi)- (a^{3/2}\dot\phi) \sin(\mu\phi).\label{Py}
\eqn
Because there are no dynamics in the lapse function, without loss of generality we choose $N(t)=1$. Then, we find that 
\bqn
 P_a \equiv \frac{\partial \mathcal{L}}{\partial \dot{a}}= -6\dot{a}a,\quad
P_\phi \equiv \frac{\partial \mathcal{L}}{\partial \dot{\phi}}= -a^3\dot{\phi}.
\eqn
Substituting $P_a$ and $P_\phi$ into Eq.(\ref{Px}) we obtain 
\begin{equation}
P_{0x}=-\frac{P_a}{4a^{1/2}\mu}\sin(\mu\phi)-\frac{P_\phi}{a^{3/2}}\cos(\mu\phi),\end{equation}
\begin{equation}
P_{0y}=-\frac{P_a}{4a^{1/2}\mu}\cos(\mu\phi)+\frac{P_\phi}{a^{3/2}}\sin(\mu\phi).
\end{equation}\\
Then, applying $P_{0x}$ and $P_{0y}$ in the momentum-deformed Hamiltonian due to GUP correction in Eq.(\ref{HgupFree}), we get
\begin{widetext}
\begin{equation}
\mathcal{H}_{GUP}= -N\left[\frac{\omega^2}{2}a^3+\frac{P_a^2}{12a}+\frac{P_\phi^2}{2a^3}-\beta\gamma\left(\frac{P_a^2}{6a}+\frac{P_{\phi}^2}{a^3}\right)^{3/2} +\gamma^2\left(\frac{P_a^2}{6a}+\frac{P_{\phi}^2}{a^3}\right)^2\left(\frac{\beta^2}{6}+\frac{2\epsilon}{3}\right)+ \mathcal{O}(\gamma^3)\right]\label{beta},
\end{equation}
  Now, the full dynamics as dictated by the Hamiltonian of Eq. (\ref{beta})  involve all the three parameters $\beta$, $\epsilon$, and $\gamma$. To simplify the analysis and obtain physically relevant dynamics, we focus on the $\gamma$-axis of the quantum parameter phase space, although  the three quantum parameters $\beta$, $\epsilon$, and $\gamma$ are in principle independent. So, the GUP-modified Hamiltonian is 
 \begin{equation}
\scalebox{1.1}{$\mathcal{H}_{GUP}=-N\left[\frac{\omega^2}{2}a^3+\frac{P_a^2}{12a}+\frac{P_\phi^2}{2a^3}+ 2\gamma^2\left(\frac{P_a^4}{108a^2}+\frac{P_\phi^4}{3a^6}+\frac{P_a^2P_\phi^2}{9a^4}\right)\right]$}.\label{m}
\end{equation}
This is the required GUP distorted Hamiltonian in the cosmological phase space dominated by a phantom scalar field with a positive cosmological constant up to second-order perturbation.  The corresponding  Raychaudhuri equation reads
\begin{equation}
2\frac{\Ddot{a}}{a}+\left(\frac{\dot a}{a}\right)^2 =\frac{\dot\phi^2}{2}-\frac{\omega^2}{2}+\gamma^2\left(16a^3 H ^4 +\frac{4a^3\dot\phi^4}{3}+\frac{32a^3 H^2\dot\phi^2}{3}\right)\label{rayforlambda},
\end{equation}
while the modified Friedmann equation is given by
\begin{equation}
3H^2=-\left(\frac{\dot\phi^2}{2}+\frac{\omega^2}{2}\right)-2\gamma^2\left(12H^4a^3+\frac{a^3\dot\phi^4}{3}+4a^3H^2\dot\phi^2\right)\label{freidmann}.
\end{equation}
\end{widetext}
It is interesting to note that and the KG equation is
\begin{equation}
\Ddot{\phi}+3\dot\phi H=0 \label{klein}.
\end{equation}
We notice that Eqs.(\ref{rayforlambda}) and  (\ref{freidmann}) directly incorporate quantum corrections, whereas for the Klein-Gordon equation (\ref{klein}),  there is no explicit dependence on quantum corrections. Any quantum deformation entering Eq.(\ref{klein}) arises only implicitly through the Hubble parameter. This situation is similar to the models of LQC \cite{rn,sn,sn1,sn2,sn3,b1,ashtekar2015generalized,ashtekar2007loop}. In this paper, we consider only the quadratic momentum deformed GUP and use the modified Hamiltonian in Eq. (\ref{m}) to study the deformed dynamics of cosmology with a cosmological constant and an arbitrary potential.

\subsubsection{GUP corrected Friedmann equation with cosmological constant}\label{freidmanndsa}

In this subsection we simplify the Friedmann Eq.(\ref{freidmann}).
Defining a new parameter $\alpha  \equiv 2\gamma ^2$,  we find that Eq.(\ref{freidmann}) can be written as a quadratic equation in terms of  $\Tilde{H}=H^2$ as
\begin{equation}
3\Tilde{H}^2 +\left(\dot{\phi}^2+\frac{3}{4\alpha a^3}\right)\Tilde{H}-C_0=0,
\end{equation}
where 
\bqn
 C\equiv-\left(\frac{\dot\phi^2}{2}+\frac{\omega^2}{2}+\frac{\alpha a^3\dot\phi^4}{3}\right),\quad
 C_0\equiv\frac{C}{4\alpha a^3}. ~~~~
\eqn
Solving the quadratic equation gives us
\begin{equation}
\Tilde{H}=\frac{-\left(\dot{\phi^2}+\frac{3}{4\alpha a^3}\right) \pm \sqrt{\left(\dot \phi^2+\frac{3}{4\alpha a^3}\right)^2+12C_0}}{6}\label{real}.
\end{equation}
Since $\Tilde{H}=H^2$, the right-hand side of the Eq.(\ref{real}) must be greater than zero. This implies
\begin{equation}
\sqrt{\left(\dot \phi^2+\frac{3}{4\alpha a^3}\right)^2+12C_0}  > \left( \dot \phi^2+\frac{3}{4\alpha a^3}\right)\label{condition},
\end{equation}
which is equivalent to
\begin{equation}
C_0>0,
\end{equation}
or
\begin{equation}
\frac{\dot \phi ^2}{2}+\frac{\alpha a^3 \dot \phi^4}{3} < \frac{3\Lambda}{8}. \label{lambdacondition}
\end{equation}
\begin{widetext}
Rewriting Eq.(\ref{real})
\begin{equation}
H^2=-\frac{\dot \phi^2}{6}-\frac{1}{8\alpha a^3}+\left(\frac{\dot \phi ^2}{6}+\frac{1}{8\alpha a^3}\right) \sqrt{1+\frac{12C_0}{\left(\dot \phi ^2+\frac{3}{4\alpha a^3}\right)^2}}
\label{rooteq}      ,
\end{equation}
and applying the conditions Eq.(\ref{condition}), where $M\equiv\dot \phi ^2 +\frac{3}{4\alpha a^3}$, we have
\begin{equation}
\sqrt{1+\frac{12C_0}{M^2}} > 1 .
\end{equation}
and 
\begin{equation}
 \quad  \frac{12 C_0}{M^2}= \frac{6\dot \phi^2 -\frac{9}{2}\Lambda +4\alpha a^3 \dot \phi^4}{6 \dot \phi^2+\frac{9}{4 \alpha a^3}+4 \alpha a^3 \dot \phi^4}<<1. \label{condi}
\end{equation}
Comparing the numerator and denominator of the Eq.(\ref{condi}), we find that the first term and the last term are the same, while in the second term, the denominator has a $\alpha^{-1}$-dependence. Since $\alpha$, which represents the GUP correction, is taken to be very small and during the early epoch the scale factor $a(t)$ is also very small, which makes the denominator very large. 

Applying binomial expansion of the Eq.(\ref{rooteq}) we get,
\begin{equation}
H^2=-\frac{\dot \phi^2}{6}-\frac{1}{8\alpha a^3}+\left(\frac{\dot \phi ^2}{6}+\frac{1}{8\alpha a^3}\right)\left [1+\frac{1}{2}\left(\frac{12C_0}{M^2}\right)-\frac{1}{2}\left(\frac{1}{2}-1\right)\left(\frac{1}{2!}\right)\left(\frac{12C_0}{M^2}\right)^2+ \mathcal{O}(3)\right] .
\end{equation}
\end{widetext}
Considering term up to the first order in $12C_0/M^2$, we find
\begin{equation}
H^2=\frac{1}{4\alpha a^3\dot\phi ^2 + 3}\left(\frac{-\dot \phi^2}{2}+\frac{3\Lambda}{8}-\frac{\alpha a^3 \dot \phi^4}{3}\right), \label{finalfreidforlambda}
\end{equation}
which   is the required Friedmann equation with GUP modification for a phantom scalar field with a positive cosmological constant.

\subsection{Phantom scalar field with an arbitrary potential
} \label{potential}

Now, we construct the GUP-modified Friedmann equation for arbitrary potential by applying the change in variables shown in the previous section.
 The Lagrangian can be written in terms of ($x$,$y$,$\dot x$,$\dot y$) by applying the procedure  prescribed in \ref{GUP deformed}, so we find 
\begin{equation}
\mathcal{L}= -\left[\frac{\dot x^2 + \dot y^2}{2} + \frac{3}{8}(x^2+y^2)V(\phi)\right] .\label{Langrange}
\end{equation}
In addition, the unperturbed Hamiltonian can be obtained by the Legendre transformation of Eq.(\ref{Langrange})
\begin{equation}
H_0 = -\left[\frac{P_x^2}{2}+\frac{P_y^2}{2}-\frac{3}{8}(x^2+y^2)V(\phi)\right] .\label{original Hamil}
\end{equation}
\begin{widetext}
Introducing GUP, as given in section \ref{GUP deformed} to the unperturbed Hamiltonian Eq.(\ref{original Hamil}), we find 
\begin{equation}
\mathcal{H}_{GUP}=-N\left[\frac{P_a^2}{12a}+\frac{P_\phi^2}{2a^3} -a^3V(\phi)+ 2\gamma^2\left(\frac{P_a^4}{108a^2}+\frac{P_\phi^4}{3a^6}+\frac{P_a^2P_\phi^2}{9a^4}\right)\right] . \label{GUP Hamil}
\end{equation}
From the above GUP corrected Hamiltonian  one can easily obtain the Raychaudhuri equation, given by
\begin{equation}
2\frac{\Ddot{a}}{a}+\left(\frac{\dot a}{a}\right)^2 =\frac{\dot\phi^2}{2}+V(\phi)+\gamma^2\left(16a^3 H ^4 +\frac{4a^3\dot\phi^4}{3}+\frac{32a^3 H^2\dot\phi^2}{3}\right)\label{ray for potential} .
\end{equation}
\end{widetext}
Following the same procedure given in section \ref{freidmanndsa}, after some algebraic manipulations, the Friedmann equation for an arbitrary potential can be written as
\begin{equation}
H^2=\frac{1}{4\alpha a^3\dot\phi ^2 + 3}\left(-\frac{\dot \phi^2}{2}+ V(\phi)-\frac{\alpha a^3 \dot \phi^4}{3}\right), \label{finalfreid for potential}
\end{equation}
and the KG equation as 
\begin{equation}
\Ddot{\phi}+3\Dot{\phi}H-\frac{dV(\phi)}{d\phi}=0.
\end{equation}

{With the same reasons as explained previously, now the KG equation also remains the same.}
We note that despite the inclusion of the potential term there is no explicit quantum correction to the above KG equation. In addition to the reason explained in Sec.\ref{GUP deformed} concerning GUP correction only for kinetic part of the phantom scalar field, however, in the present section, the potential is treated classically in the light of the models of LQC \cite{rn,sn,sn1,sn2,sn3,b1,ashtekar2015generalized,ashtekar2007loop} The justification of this treatment is based on the fact that as the potential begins to dominate on the onset inflationary era the quantum effects of spacetime starts diluting. Thus it validates the employment of effective dynamics.

In SMC, the factor $\alpha a^3 \dot \phi^4/3$ is absent in the numerator of Eq.(\ref{finalfreid for potential}), retaining only the energy density term. The additional terms arise solely from the quantum corrections due to GUP.
In addition, Eq.(\ref{finalfreid for potential}) reveals the possibility of singularity resolution. The occurrence of non-singular bounce in phantom models has already been studied in \cite{brown2008phantom,zhu2017universal,sharma2023nonsingular,battefeld2015critical,cai2012towards}. The possibility of a non-singular bounce is further enhanced by the presence of a negative quantum corrected term, $-\alpha a^3 \dot \phi^4/3$. Though $\dot a =0$ is a necessary condition for the occurrence of a bounce, the solution must meet $\ddot a>0$ at the   bounce for the contracting universe to reverse its trajectory and begin expanding.
This condition is satisfied by the Raychaudhuri equation (\ref{ray for potential}),  as its right-hand side is positive.

\section{
Dynamical System Analysis
}\label{DSA}
A nonlinear system is generally difficult to study analytically. However, the method of dynamical system analysis (DSA) serves as a powerful tool for extracting qualitative information. Einstein's equations, when applied to the flat FLRW spacetime, become a set of coupled second-order ordinary differential equations. However, with a suitable choice of variables, they can be transformed to the first-order autonomous differential equations.

In this method the system's dynamics is cast as a set of first-order autonomous differential equations, and the fixed points are defined as the points where the vector flow of the dynamical variables vanishes. The precise nature of the fixed points is obtained by examining the behavior of the leading order perturbation around the fixed points. Mathematically, the signs of the eigenvalues of the Jacobian matrix evaluated at the fixed points indicate the nature of the fixed points.  For an extensive review of DSA, we refer readers to \cite{alho2022cosmological,s17,shahalam2015dynamics,Jaskirat}.

Phase portraits, on the other hand, are visual representations of the trajectories of a dynamic system. It provides insights into the qualitative behavior of the system pictorially. 
This is achieved by drawing a tangent at each point given by the flow vectors of the autonomous differential equations. When applied to cosmology, it offers an intuitive understanding of the fate of the Universe even without solving the equations.

In the subsequent subsections, we construct the autonomous equations and henceforth perform DSA for the background  with the GUP modification introduced in the above sections.

We know that  Einstein's equations are second order in nature. However, to perform a DSA, they must be transformed into a set of first-order differential equations. A widely practiced method is to begin by normalizing the Friedmann equation by the square of the Hubble parameter to make each term dimensionless. In the process, it brings all the components contributing to the Hubble rate on equal footing. The next step is to write the EoM for each independent dimensionless dynamical variable obtained with the help of the Raychaudhuri and Klien-Gordon equations. The final set of the equations, when expressed entirely in terms of newly defined dimensionless variables, constitutes the required autonomous system. 

In the following, let us consider some dynamical cases with different potentials $V(\phi)$, separately. 
 
\subsection{For $V(\phi)=V_0 \phi^2$ }\label{square potential}

Following the above-mentioned recipe, we perform DSA by constructing autonomous equations for the chosen potential. This will allow us to perform fixed point analysis and study the behavior of phase portraits.
From  Eq.(\ref{finalfreid for potential}),  the expansion normalized Friedmann equation can be written as
\begin{equation}
4 \alpha a^3 \dot \phi^2 +3 =-\frac{\dot \phi^2}{2H^2}+\frac{V(\phi)}{H^2}-\frac{2 \alpha a^3 \dot \phi^2}{3}\frac{\dot \phi^2}{2H^2}.
\end{equation}
A suitable choice of dimensionless dynamical variables,  also called the expansion normalised  (EN) variables, are 
\begin{eqnarray}\label{1}
x &\equiv& \sqrt{\frac{\dot \phi ^2}{6 H^2}}, \quad 
y \equiv \sqrt{\frac{V_0\phi^2}{3H^2}}, \quad 
z \equiv \sqrt{\alpha a^3 \dot \phi^2}. 
\end{eqnarray}
Then, 
the expansion-normalized Friedmann equation reads
\begin{eqnarray}\label{2}
x^2 + y^2-\frac{2}{3} z^2 x^2-\frac{4}{3}z^2=1. \label{constr V}
\end{eqnarray}
{It is observed that the GUP corrections denoted by $\alpha$  is presented only in the variable $z$ in Eq.}(\ref{2}) as can be seen from Eq.(\ref{1}). In addition,
 from Eq.(\ref{Vklein}) we find that  the KG equation reads
\begin{equation}
\ddot{\phi}+3\dot{\phi}H-2V_0\phi=0.
\end{equation}
From Eq.(\ref{ray for potential}), on the other hand,  we  obtain
\begin{eqnarray}
\frac{\dot{H}}{H^2}&=&\frac{3}{2}(x^2+y^2-1)\nonumber\\
&& +\frac{(2+6x^4+8x^2)(-1-x^2+y^2)}{(2x^4+4x^2)}.
\end{eqnarray}

It should be noted that the EN variables alone fail to close the autonomous system for the power law potential. For example, a new dynamic variable depending on $\phi$ appears in the form of $\lambda\equiv -\frac{V,_\phi}{V}$. 
The physical phase space for a power law system is always represented by a positive $y$ half-cylinder stretching from $\lambda=0 $ to $+\infty$ due to the symmetry \cite{gong2014general,b18,ng2001applications,r14}, which means that the phase space is not compact. To make phase space compact, we choose a new dynamic variable $u$, which is
\begin{equation}
u=\frac{\lambda}{\lambda+1}.
\end{equation}
This transformation makes our phase space compact with the range $0\leq u\leq 1$. finally we can write our system of equations in terms of $u$.
Then, the autonomous set of dynamical equations for quadratic potential with the GUP correction finally read 
\begin{widetext}
\begin{eqnarray}
f(x,y) &\equiv& \frac{dx}{d N}= -3(1-u)x -u\left(\sqrt{6}y^2 \right)- (1-u)x \left\{\frac{3}{2}(x^2+y^2-1)+\frac{(2+6x^4+8x^2)(-1-x^2+y^2)}{(2x^4+4x^2)}\right\}, \label{square 1} \\
g(x,y) &\equiv& \frac{dy}{dN}=\sqrt{6}yxu -y(1-u)\left\{\frac{3}{2}(x^2+y^2-1)+\frac{(2+6x^4+8x^2)(-1-x^2+y^2)}{(2x^4+4x^2)}\right\}, \label{square2}\\
h(x,y) &\equiv& \frac{d u}{d N}=- \sqrt{6}(\Gamma-1) (1-u) x z^2, \quad \Gamma= \frac{V V_{,\phi \phi}}{V_{,\phi}^2}.
\label{square3}
\end{eqnarray}
The  cosmological dynamics for the universe with a quadratic potential and the GUP corrections are completely contained in the three equations (\ref{square 1}), (\ref{square2}) and (\ref{square3}). We also observe that the GUP corrections appear only in the third autonomous equation,  for $h(x,y)$.
\end{widetext}



Now, we perform a thorough analysis of the fixed points of the cosmological system dictated by a phantom scalar field with quadratic potential in a GUP-modified scenario. Later, we compare the results with those without GUP.

Fixed points are obtained by setting $d {\cal{F}}/dN = 0$ simultaneously, where ${\cal{F}} \equiv (x, y, u)$. This physically means that the system becomes stationary at the points. 
Then, the fixed points along with their behavior are tabulated in Table \ref{Table2}. To compare our results with the original dynamics, we turn off the quantum perturbation by setting $\alpha=0$ in Eqs.(\ref{square 1}-\ref{square3}), and the  results are summarized in Table \ref{Table1}.

Comparison of Tables \ref{Table1}
and \ref{Table2} reveals the dynamics without GUP are richer than the dynamics after GUP modifications are taken into account. This is clear as Table \ref{Table1}   contains a greater number of fixed points than Table \ref{Table2} with the GUP corrections. However, introducing distortion due to the GUP effects does not change  the fixed points as the eigenvalues in both tables contain zero.

\begin{widetext}

\begin{table}[htbp]
\centering
\caption{\textbf{Fixed points and the stability analysis for the Potential $V(\phi)=V_0\phi^2$ without GUP}}
\resizebox{0.6\linewidth}{!}{
\begin{tabular}{l c c c c c c c}
\toprule
& \textbf{x} & \textbf{y} & $u$ & \textbf{$E_1$} & \textbf{$E_2$}  & \textbf{$E_3$} & \textbf{Stability}\\
\midrule
\textbf{A} & 0 & 0 & c & $-3(1-c)/2$ & $3(1-c)/2$& 0 & saddle point \\
\textbf{B} & 0 & 0 & 0 & $-3/2$ & $3/2$& 0 & saddle point \\
\textbf{C} & 0 & 0 & 1 & 0 & 0 & 0 & neutral point \\
\textbf{D} & $c$ & 0 & 1 & $2\sqrt{6}c$ & $-\sqrt{6}c$& 0 & saddle point \\
\textbf{E} & 0 & -1 & 0 & $-3$ & $-3$& 0 & stable point \\
\textbf{F} & 0 & 1 & 0 & $-3$ & $-3$& 0 & stable point \\
\bottomrule
\end{tabular}
}
\label{Table1}

\end{table}
\begin{table}[htbp]
\centering
\caption{\textbf{Fixed points and the stability analysis for the Potential $V(\phi)=V_0\phi^2$} with GUP}
\resizebox{0.6\linewidth}{!}{
\begin{tabular}{l c c c c c c c}
\toprule
& \textbf{x} & \textbf{y} & $u$ & \textbf{$E_1$} & \textbf{$E_2$}  & \textbf{$E_3$} & \textbf{Stability}\\
\midrule
\textbf{A} & $+c$(other than 0) & 0 & 1 & 0 & $2\sqrt{6}c$& $-\left(\frac{2\sqrt{6}c^3+\sqrt{6}c^5}{c^2(2+c^2)}\right)$ & saddle point \\
\textbf{B} & $-c$(other than 0) & 0 & 1 & 0 & $ -2\sqrt{6}c$& $\left(\frac{2\sqrt{6}c^3+\sqrt{6}c^5}{c^2(2+c^2)}\right)$ & saddle point \\

\bottomrule
\end{tabular}
}
\label{Table2}

\end{table}
\end{widetext}

\subsection{Exponential Potential}

In this subsection, we consider a potential of the form 
\begin{equation}
V(\phi)=V_0e^{-k\phi},
\end{equation}
to study the GUP modified background dynamics, where $V_0$ and $k$ are constants. Then, the EN variables are 
\begin{eqnarray}
x \equiv \sqrt{\frac{\dot \phi ^2}{6 H^2}}, \quad 
y \equiv \sqrt{\frac{V_0e^{-k\phi}}{3H^2}}, \quad 
z \equiv \sqrt{\alpha a^3 \dot \phi^2}. ~~~~ 
\end{eqnarray}
it cna be shown that now the expansion-normalized Friedmann equation in the form of the EN variables reads
\begin{eqnarray}
x^2 + y^2-\frac{2}{3} z^2 x^2-\frac{4}{3}z^2 =1,
\end{eqnarray}
and from Eq.(\ref{Vklein}) the Klein Gordon equation  reads 
\begin{equation}
\ddot{\phi}+3\dot{\phi}H+V_0ke^{-k\phi}=0.
\end{equation}
Furthermore, the Raychaudhuri equation (\ref{ray for potential}) in terms of the EN variables is given by
\begin{eqnarray}
 \frac{\dot{H}}{H^2} &=& \frac{3}{2}(x^2+y^2-1)\nonumber\\
&& +\frac{(2+6x^4+8x^2)(-1-x^2+y^2)}{(2x^4+4x^2)}.
\end{eqnarray}
Because our potential is exponential, we can get  $\Gamma= \frac{V V_{,\phi \phi}}{V_{,\phi}^2}$ equal to 1. While constructing the autonomous equation, we obtain a factor of $Q\equiv-\frac{V_{,\phi}}{V}= k$.
With these, we can write the autonomous differential equations as
\begin{widetext}
\begin{eqnarray}
\Tilde{f}(x,y)&=&\frac{dx}{d N}= -3x - \sqrt{\frac{3}{2}} Q y^2 - x \left\{\frac{3}{2}(x^2+y^2-1)+\frac{(2+6x^4+8x^2)(-1-x^2+y^2)}{(2x^4+4x^2)}\right\}, \label{first}\\
\Tilde{g}(x,y)&=&\frac{dy}{dN}=-\sqrt{\frac{3}{2}} Q yx -y\left\{\frac{3}{2}(x^2+y^2-1)+\frac{(2+6x^4+8x^2)(-1-x^2+y^2)}{(2x^4+4x^2)}\right\}.\label{sec}
\end{eqnarray}
\end{widetext}
The original dynamics without the quantum corrections  can be easily obtained as a limiting case by setting $\alpha=0$.

\subsubsection{Fixed pointS}

The fixed points of the system with the eigenvalues of the Jacobian are summarized in  Tables \ref{Table3} and \ref{Table4}, respectively, with and without GUP.
\begin{widetext}

\begin{table}[htbp]
\centering
\caption{\textbf{Fixed points and corresponding eigenvalues of the Potential $V(\phi)= V_0 e^{-k\phi} $} without GUP}
\resizebox{0.7\linewidth}{!}{
\begin{tabular}{l c c c c c c}
\toprule
& \textbf{x} & \textbf{y}  & \textbf{$E_1$} & \textbf{$E_2$}  & \textbf{Stability}\\
\midrule
\textbf{A} &  $-Q/\sqrt{6}$& $\sqrt{(Q^2+6)/6}$& $-(6+Q^2)/2 $ & $-(3+Q^2)$ & stable point\\
\textbf{B} & $-Q/\sqrt{6}$& $-\sqrt{(Q^2+6)/6}$  & $-(6+Q^2)/2 $ & $-(3+Q^2)$ & stable point\\
\textbf{C} & 0 & 0 & $-3/2$ & $3/2$ & saddle point\\
\bottomrule
\end{tabular}
}
\label{Table3}
\end{table}
\begin{table}[htbp]
\centering
\caption{\textbf{Fixed points and the corresponding eigenvalues of the Potential $V(\phi)= V_0 e^{-k\phi} $} with GUP.}
\resizebox{0.8\linewidth}{!}{
\begin{tabular}{l c c c c c c}
\toprule
& \textbf{x} & \textbf{y} &  \textbf{$E_1$} & \textbf{$E_2$}  & \textbf{Stability}\\
\midrule
\textbf{A} & $-\frac{Q}{\sqrt{6}}  $& $\sqrt{Q^2+6/6}$   & $\frac{-72Q^2-18Q^4-Q^6}{2Q^2(12+Q^2))}$ & $-\frac{72+84Q^2+21Q^4+Q^6}{Q^2(12+Q^2))}$ & stable point\\
\textbf{B} & $-\frac{Q}{\sqrt{6}} $& $-\sqrt{Q^2+6/6}$  & $\frac{-72Q^2-18Q^4-Q^6}{2Q^2(12+Q^2))}$ & $-\frac{72+84Q^2+21Q^4+Q^6}{Q^2(12+Q^2))}$ & stable point\\
\bottomrule
\end{tabular}
}
\label{Table4}
\end{table}
\end{widetext}
The rigorous fixed point analysis shows that the introduction of GUP corrections completely alters the fixed points and hence their stability in the case of exponential potential. It is found that the dynamics without GUP have two stable fixed points and one saddle point, whereas the GUP-modified dynamics for the exponential potential leave us only with two stable fixed points.

\subsubsection{Phase portrait}

\begin{widetext}

\begin{figure}[htbp]
\centering
\begin{minipage}[t]{0.45\textwidth}
\centering
\includegraphics[width=0.8\textwidth]{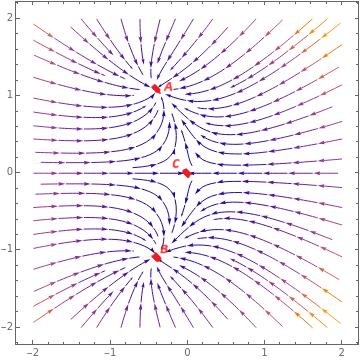}
\end{minipage}
\hfill
\begin{minipage}[b]{0.45\textwidth}
\centering
\includegraphics[width=0.8\textwidth]{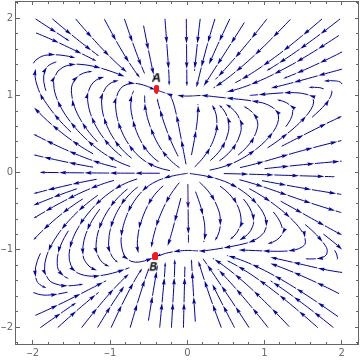}
\end{minipage}
\caption{{Phase portrait for potential $V(\phi)= V_0 e^{-k\phi} $ for value of $Q=1$. \textit{Left} without GUP fluctuation where the stable fixed points are at \textbf{A}=(-0.40,1.08), \textbf{B}=(-0.40,-1.08)  and a saddle fixed point at \textbf{C}=(0,0)} and \textit{Right} with GUP where the stable fixed point are \textbf{A}=(-0.40,1.08), \textbf{B}=(-0.40,-1.08) }\label{Q1}
\end{figure}
\end{widetext}

However, from the physical perspective only the upper half of the plane $y\geq 0$ is of meaning. This is because $y\leq0$ is not physically achievable for a positive potential. Thus, we discard the set of the points, including the fixed points, with $y\leq 0$. The dynamic variable $x$ is proportional to the velocity. In both cases, with and without GUP, the phase portraits of the system are presented in  Fig.(\ref{Q1}) for $Q =  1$. the saddle point $C=(0,0)$ disappears upon the introduction of the GUP distortion. However, the physically relevant point $A=(-0.40, 1.08)$ remains even after quantum corrections are taken into account. Physically, the stable fixed point $A=(-0.40,1.08)$ implies that at a later time, the scalar field settles down to a negative velocity value with a positive field value.

\section{
GUP-modified inflationary scenario
} \label{inflation}

In this section, we present inflationary dynamics under the influence of GUP for quadratic and exponential potentials. To understand the change in the behavior of the background dynamics of our cosmology in the presence of GUP distortions, we need to study the behavior of cosmologically relevant parameters, such as, the expansion factor $a(t)$, the scalar field $\phi(t)$,  the EoS ($w_{eff}$), and the slow-climb parameters $\epsilon$ and $\eta$ with and without GUP deformations \cite{piao2004phantom}. In what follows, we do this case by case for each potential considered in this article.

\subsection{Quadratic Potential $V(\phi)=\mu^2 \phi^2$} \label{quadratic}

During cosmic inflation, the behavior of a phantom field differs from that of a normal scalar field. While a normal scalar field undergoes a slow roll along its potential, a phantom field exhibits a slow climb along its potential. This distinction is evident both mathematically and graphically, as demonstrated in \cite{sami2004phantom}  for power-law potentials.

After introducing GUP corrections to the Friedmann equations, the slow climb parameters defined as  
\bq
\epsilon\equiv-\frac{\dot H}{H^2}, \quad \eta\equiv\frac{V"}{3H^2},
\eq
and $\delta \equiv\eta -\epsilon$ can be expressed as follows
\begin{widetext}
\begin{equation}
\epsilon = \frac{3}{2}\left(\frac{\dot \phi^2}{6H^2}+\frac{V_0\phi^2}{3H^2}-1 \right)
+4 \alpha a^3H^2+\frac{\alpha a^3\dot \phi^4}{3H^2}+\frac{8}{3}\alpha a^3\dot \phi ^2, \quad \eta = \frac{2 V_0}{3H^2},
\end{equation}
for $V(\phi)=V_0\phi^2$.
\begin{figure}[htbp]
\centering
\begin{minipage}[t]{0.45\textwidth}
\centering
\includegraphics[width=1.1\textwidth]{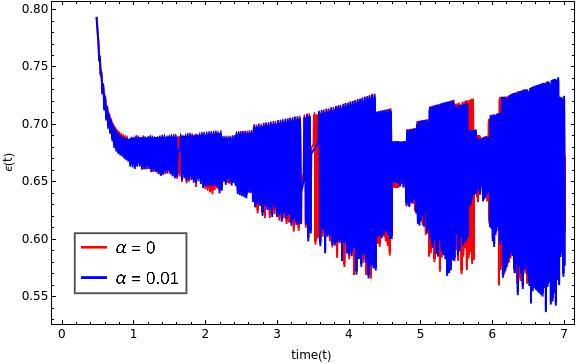}
\end{minipage}
\hfill
\begin{minipage}[b]{0.45\textwidth}
\centering
\includegraphics[width=1.09\textwidth]{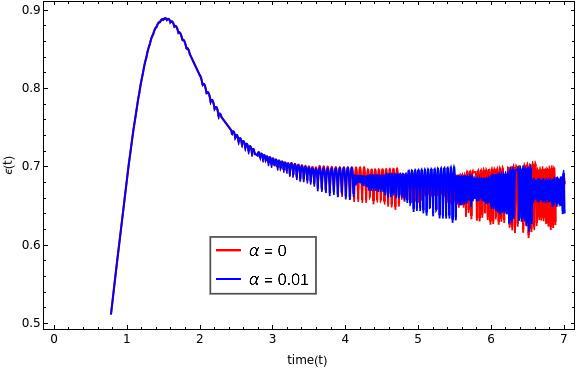}
\end{minipage}
\caption{\textit{Left panel}: Plot of $\epsilon(t)$ vs time(t) for the initial condition as $a(0)=1$, $\dot a(0)=1$ and $\dot \phi(0)=0.1$ for three different values of $\alpha$ where, $\alpha =0 $ shows no GUP fluctuation. \textit{Right panel}: For initial condition as $a(0)=4$, $\dot a(0)=4$ and $\dot \phi(0)=0.1$.}\label{epsilon1}
\end{figure}
\begin{figure}[H]
\centering
\begin{minipage}[t]{0.45\textwidth}
\centering
\includegraphics[width=1.1\textwidth]{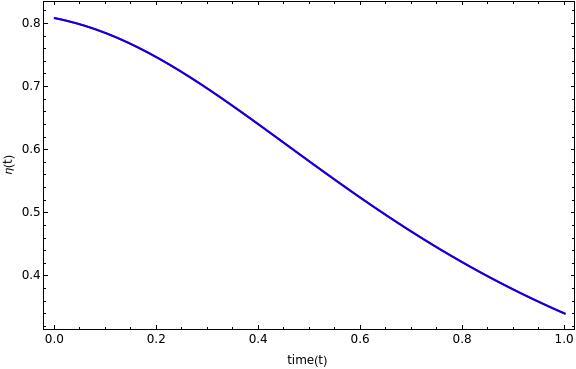}
\end{minipage}
\caption{$\eta(t)$ vs $t$ for the square potential. }
\label{ etasquare}
\end{figure}
\end{widetext}
In the cosmological context, EoS is a useful parameter to understand and classify the acceleration and deceleration phases of our universe. 
For $w=0$, it corresponds to non-relativistic matter such as cold dark matter (CDM) or non-relativistic baryonic matter and for $w = 1/3$ it refers to radiation dominated. For $w=-1$,$-1<w<-1/3$ and $w<-1$ refer to the cosmological constant, Quintessence, and Phantom eras, respectively.

The Raychaudhuri equation in SMC can be written in terms of EoS as follows
\begin{equation}\frac{\ddot a}{a}=-(1+3 w)\frac{\rho}{6},
\end{equation}
while the GUP modified Raychaudhuri Eq.(\ref{ray for potential}) reads
\begin{equation}
\frac{\ddot a}{a}=-\left[1+3\left(\frac{\frac{-\dot \phi^2}{2}-V(\phi)-\frac{2\alpha a^3 \dot \phi^4}{3}-\frac{4\alpha a^3 \dot \phi^2 V(\phi)}{3}}{-\frac{\dot \phi^2}{2}+V(\phi)-\frac{\alpha a^3 \dot \phi^4}{3}}\right)\right]\frac{\rho}{6}.\label{weq.}
\end{equation}
Hence, 
we get
\begin{equation}
w = \frac{\frac{-\dot \phi^2}{2}-V(\phi)-\frac{2\alpha a^3 \dot \phi^4}{3}-\frac{4\alpha a^3 \dot \phi^2 V(\phi)}{3}}{-\frac{\dot \phi^2}{2}+V(\phi)-\frac{\alpha a^3 \dot \phi^4}{3}}.
\end{equation}
Then, for the quadratic potential,  we have
\begin{equation}
w = \frac{\frac{-\dot \phi^2}{2}-V_0\phi^2-\frac{2\alpha a^3 \dot \phi^4}{3}-\frac{4\alpha a^3 \dot \phi^2 V_0\phi^2}{3}}{-\frac{\dot \phi^2}{2}+V_0\phi^2-\frac{\alpha a^3 \dot \phi^4}{3}}. \label{weff}
\end{equation}
\begin{widetext}
\begin{figure}[H]
\centering
\begin{minipage}[t]{0.45\textwidth}
\centering
\includegraphics[width=1.1\textwidth]{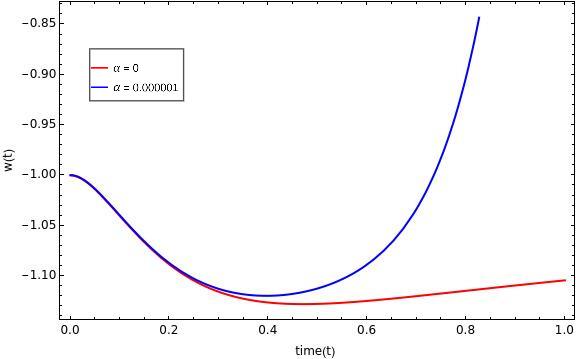}
\end{minipage}
\caption{Comparison of EoS vs time for potential $V = \mu^2\phi^2$ for different values of $\alpha$.}
\label{wsqure}
\end{figure}
\begin{figure}[H]
\centering
\begin{minipage}[t]{0.45\textwidth}
\centering
\includegraphics[width=1.1\textwidth]{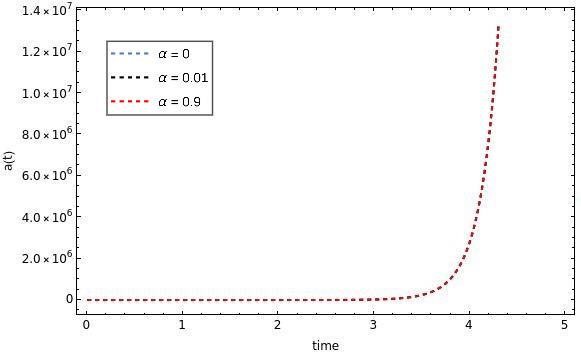}
\end{minipage}
\hfill
\begin{minipage}[b]{0.45\textwidth}
\centering
\includegraphics[width=1.09\textwidth]{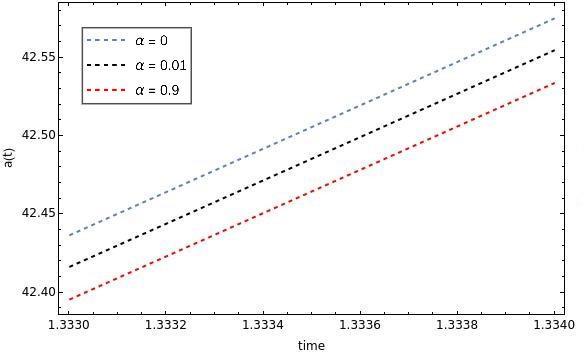}
\end{minipage}
\caption{$\textit{Left panel :}$  Plot of  $a(t)$ along the Y-axis vs time($t$) along the X-axis for three values of $\alpha=0, 0.01, 0.9$ respectively in $V = \mu^2\phi^2$ potential  with initial condition as $a(0)=1$ and $\phi(0)=1$ and $\dot\phi(0)=0.1$.
$\textit{Right panel :} $ Magnified view.}
\label{a vst 1}
\end{figure}
\begin{figure}[H]
\centering
\begin{minipage}[t]{0.45\textwidth}
\centering
\includegraphics[width=1.1\textwidth]{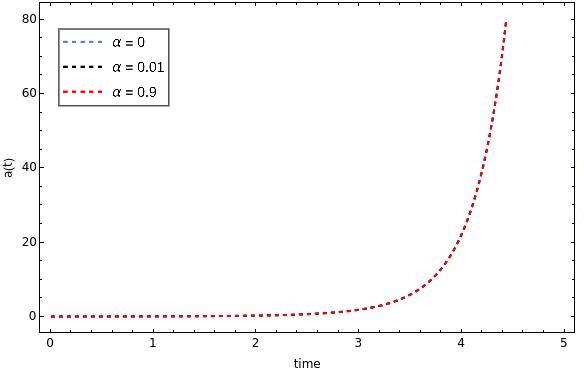}
\end{minipage}
\hfill
\begin{minipage}[b]{0.45\textwidth}
\centering
\includegraphics[width=1.16\textwidth]{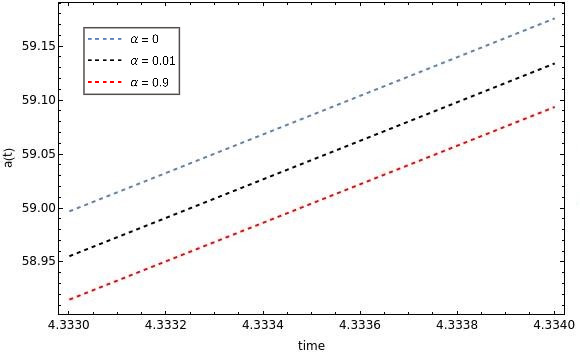}
\end{minipage}
\caption{$\textit{Left panel :}$  Plot of  $a(t)$ along the Y-axis vs time($t$) along the X-axis for three values of $\alpha=0, 0.01, 0.9$ respectively in $V = \mu^2\phi^2$ potential  with initial condition as $a(0)=0.1$ and $\phi(0)=0.1$.
$\textit{Right panel :} $ Magnified view.}
\label{a vst0.1}
\end{figure}
\begin{figure}[H]
\centering
\begin{minipage}[t]{0.45\textwidth}
\centering
\includegraphics[width=1.09\textwidth]{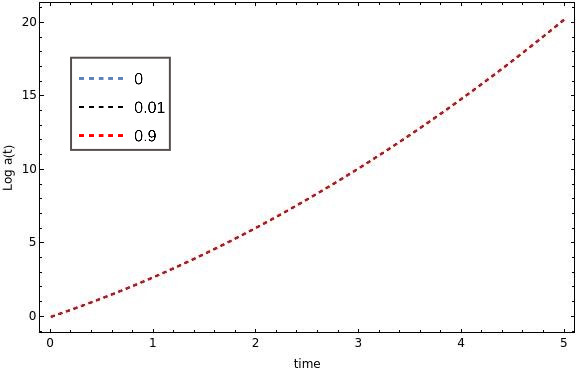}
\end{minipage}
\hfill
\begin{minipage}[b]{0.45\textwidth}
\centering
\includegraphics[width=1.14\textwidth]{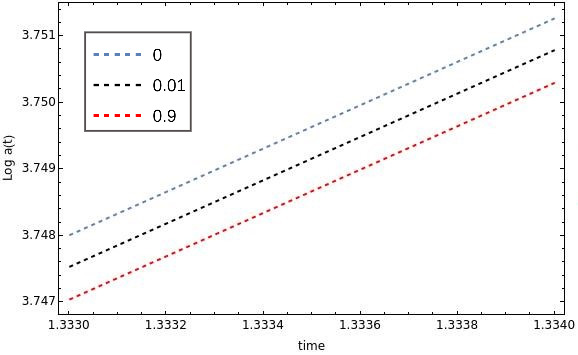}
\end{minipage}
\caption{$\textit{Left panel :}$  Plot of logarithm of $a(t)$ along the Y-axis vs time($t$) along the X-axis for three values of $\alpha=0, 0.01, 0.9$ respectively in $V = \mu^2\phi^2$ potential with initial condition as $a(0)=1$ and $\phi(0)=1$. 
$\textit{Right panel :} $ Magnified view. }\label{logavst 1}
\end{figure}
\begin{figure}[H]
\centering
\begin{minipage}[t]{0.45\textwidth}
\centering
\includegraphics[width=1.09\textwidth]{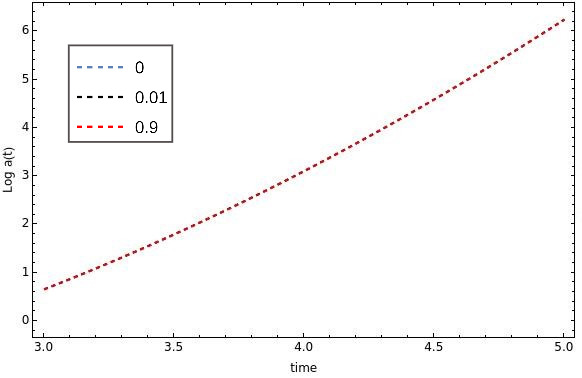}
\end{minipage}
\hfill
\begin{minipage}[b]{0.45\textwidth}
\centering
\includegraphics[width=1.15\textwidth]{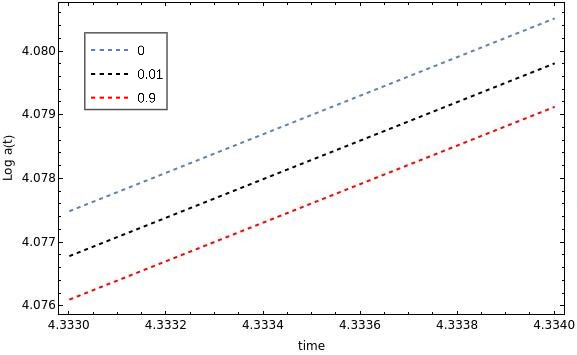}
\end{minipage}
\caption{$\textit{Left panel :}$  Plot of logarithm of $a(t)$ along the Y-axis vs time($t$) along the X-axis for three values of $\alpha=0, 0.01, 0.9$ respectively in $V = \mu^2\phi^2$ potential with initial condition as $a(0)=0.1$ and $\phi(0)=0.1$. 
$\textit{Right panel :} $ Magnified view.}\label{loga vs t 0.1}
\end{figure}
\begin{figure}[H]
\centering
\begin{minipage}[t]{0.45\textwidth}
\centering
\includegraphics[width=1.09\textwidth]{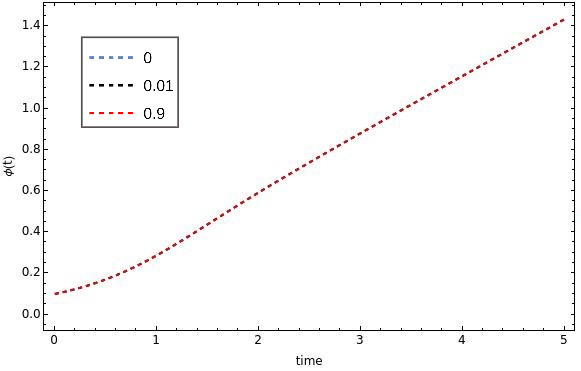}
\end{minipage}
\hfill
\begin{minipage}[b]{0.45\textwidth}
\centering
\includegraphics[width=1.15\textwidth]{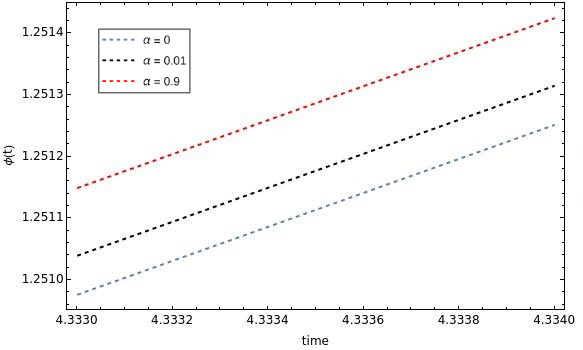}
\end{minipage}
\caption{$\textit{Left panel :}$  Plot of $\phi(t)$ along the Y-axis vs time($t$) along the X-axis for three values of $\alpha=0, 0.01, 0.9$ respectively in $V = \mu^2\phi^2$ potential with initial condition as $a(0)=1$ and $\phi(0)=1$. 
$\textit{Right panel :} $ Magnified view.}
\label{ phi1}
\end{figure}

\begin{figure}[H]
\centering
\begin{minipage}[t]{0.45\textwidth}
\centering
\includegraphics[width=1.1\textwidth]{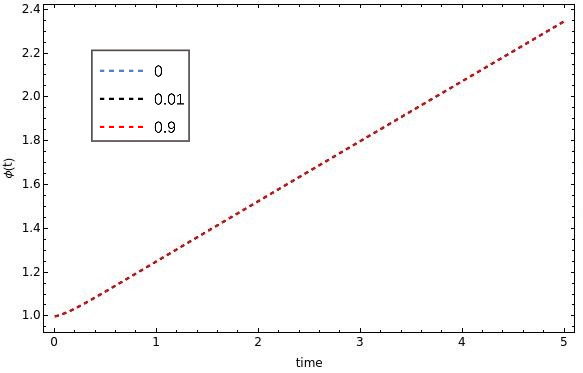}
\end{minipage}
\hfill
\begin{minipage}[b]{0.45\textwidth}
\centering
\includegraphics[width=1.15\textwidth]{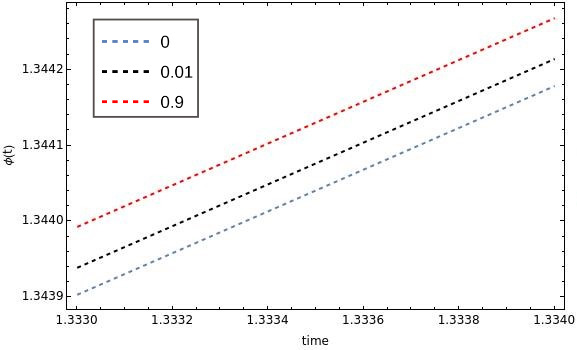}
\end{minipage}
\caption{$\textit{Left panel :}$  Plot of $\phi(t)$ along the Y-axis vs time($t$) along the X-axis for three values of $\alpha=0, 0.01, 0.9$ respectively in $V = \mu^2\phi^2$ potential with initial condition as $a(0)=0.1$ and  $\phi(0)=0.1$.
$\textit{Right panel :} $ Magnified view. }
\label{phi 0.1}
\end{figure}
\end{widetext}

Fig. \ref{a vst 1} represents the dynamics of the expansion factor, illustrating a nearly exponential rise, and hence indicating inflation. However, the effect of GUP is not readily discernible from the left side of Fig. \ref{a vst 1}. 
To observe this effect, it is necessary to magnify the range presented on the right-hand side of  Fig. \ref{a vst 1}. The blue line represents the original expansion factor without GUP modifications. As we increase the strength of  $\alpha$, the value of the original expansion factor becomes dramatically distorted.
In addition, we present the behavior of the expansion factor for different initial conditions in Fig.\ref{a vst0.1}.

The behavior of $\phi$ w.r.t the cosmic time $t$ for different initial conditions is shown in Figs. \ref{ phi1} and \ref{phi 0.1}. A magnified view of the effect of the GUP deformation for different values of $\alpha$ is provided on the right-hand side of the figures. On the left-hand sides of the figures, the scalar field starts from a very low value and then increases linearly upward for different initial conditions. As we can observe from the graph of $\phi$, the inclusion of a higher value of the GUP strength, $\alpha$ in the dynamics causes the evolution to distort upwards from the original dynamics without GUP.

Fig.\ref{wsqure} depicts the behavior of the EoS parameter with the introduction of a small strength of GUP distortions. We observe that the EoS for the phantom remains always less than $-1$ in the absence of GUP fluctuations due to the negative pressure term.
Even in this case, the behavior of the EoS remains the same for most of the evolution. Only at the later phase the EoS increases w.r.t the unperturbed case. 

In Figs. \ref{epsilon1} and \ref{ etasquare}, we present the slow climb parameters $|\epsilon|$ and $|\eta|$ as a function of time with different initial conditions and we clearly observe that the values of $|\eta|$, $|\epsilon| <<1$ indicate inflation.

\subsection{Exponential Potential  $V(\phi)= V_0e^{-k \phi}$} \label{expo.}

In this subsection, we study the GUP-modified background dynamics for exponential potential. We determine the slow roll parameters and equation of state and subsequently present them for different initial conditions. Then, the slow climb parameters are
\begin{widetext}
\begin{equation}
\epsilon = \frac{3}{2}\left(\frac{\dot \phi^2}{6H^2}+\frac{ V_0 e^{-k\phi}}{3H^2}-1 \right)+4 \alpha a^3H^2+\frac{\alpha a^3\dot \phi^4}{3H^2}+\frac{8}{3}\alpha a^3\dot \phi ^2, \quad \eta = \frac{V_0k^2 e^ {-k\phi}}{3H^2}.\\
\end{equation}
The effective equation of state (EoS) for $V(\phi)= V_0e^{-k\phi}$  is
\begin{equation}
w = \frac{\frac{-\dot \phi^2}{2}-V_0e^{-k\phi}-\frac{2\alpha a^3 \dot \phi^4}{3}-\frac{4\alpha a^3 \dot \phi^2 e^{-k\phi}}{3}}{-\frac{\dot \phi^2}{2}+V_0e^{-k\phi}-\frac{\alpha a^3 \dot \phi^4}{3}}. \label{weff}
\end{equation}

\begin{figure}[H]
\centering
\begin{minipage}[t]{0.45\textwidth}
\centering
\includegraphics[width=1.1\textwidth]{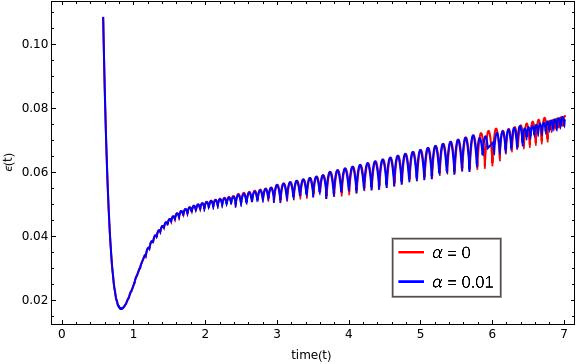}
\end{minipage}
\hfill
\begin{minipage}[b]{0.45\textwidth}
\centering
\includegraphics[width=1.09\textwidth]{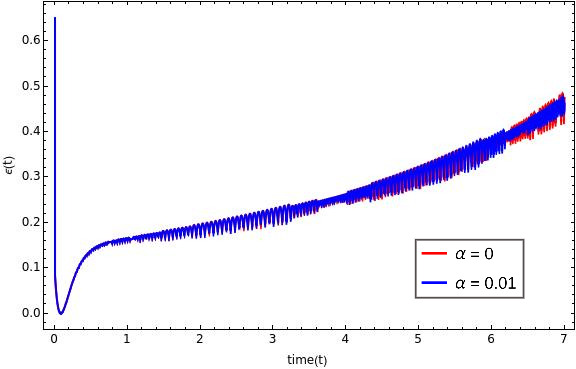}
\end{minipage}
\caption{\textit{Left panel}: For $V(\phi)= V_0e^{-k\phi}$ with initial condition as $a(0)=1$, $\dot a(0)=1$ and $\dot \phi(0)=0.1$ plot of  $\epsilon(t)$ along Y-axis vs time(t) along X-axis for three different values of $\alpha$ where, $\alpha =0 $ with no GUP fluctuation. \textit{Right panel}: For initial condition as $a(0)=0.1$, $\dot a(0)=0.1$ and $\dot \phi(0)=0.1$.}\label{epsilon2}
\end{figure}
\begin{figure}[htbp]
\centering
\begin{minipage}[t]{0.45\textwidth}
\centering
\includegraphics[width=1.1\textwidth]{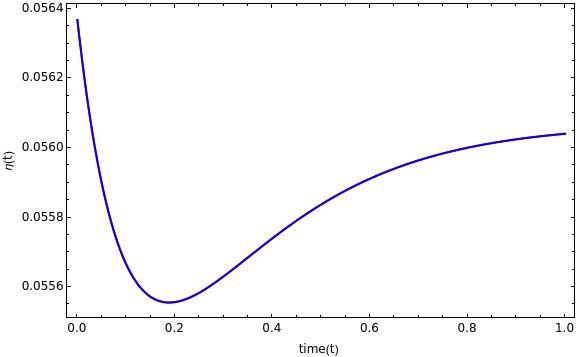}
\end{minipage}
\caption{$\eta(t)$ vs $t$ for exponential potential. }
\label{ etaexpo}
\end{figure}

\begin{figure}[H]
\centering
\begin{minipage}[t]{0.45\textwidth}
\centering
\includegraphics[width=1.1\textwidth]{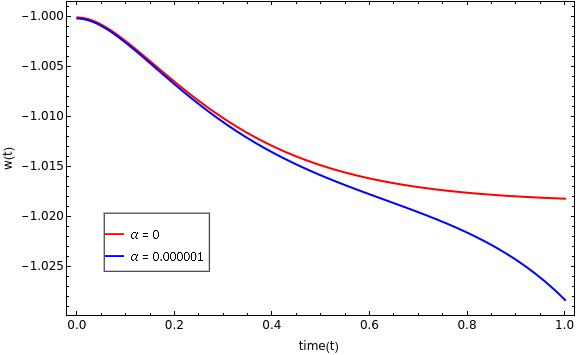}
\end{minipage}
\caption{Comparison of EoS vs time graph for potential $V = e^{-k\phi}$ for different values of $\alpha$.}
\label{ w exp.}
\end{figure}
\begin{figure}[H]
\centering
\begin{minipage}[t]{0.45\textwidth}
\centering
\includegraphics[width=1.1\textwidth]{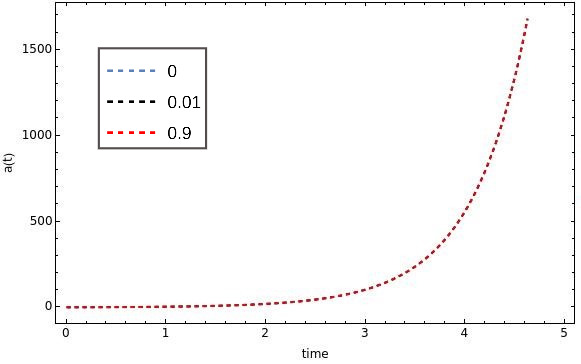}
\end{minipage}
\hfill
\begin{minipage}[b]{0.45\textwidth}
\centering
\includegraphics[width=1.13\textwidth]{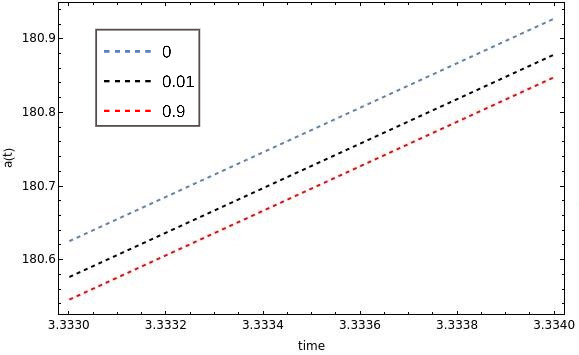}
\end{minipage}
\caption{$\textit{Left panel :}$  Plot of $a(t)$ along Y-axis vs time($t$) along X-axis for three values of $\alpha=0, 0.01, 0.9$ respectively in $V=e^{-k \phi}$ potential with initial condition as $a(0)=1$ and $\phi(0)=1$.
$\textit{Right panel :} $ Magnified view. }
\label{exp.a vs t 1}
\end{figure}
\begin{figure}[H]
\centering
\begin{minipage}[t]{0.45\textwidth}
\centering
\includegraphics[width=1.1\textwidth]{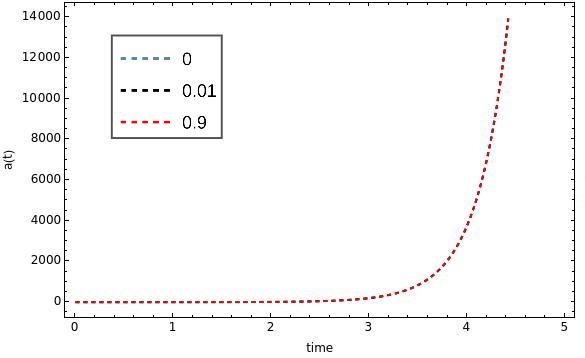}
\end{minipage}
\hfill
\begin{minipage}[b]{0.45\textwidth}
\centering
\includegraphics[width=1.1\textwidth]{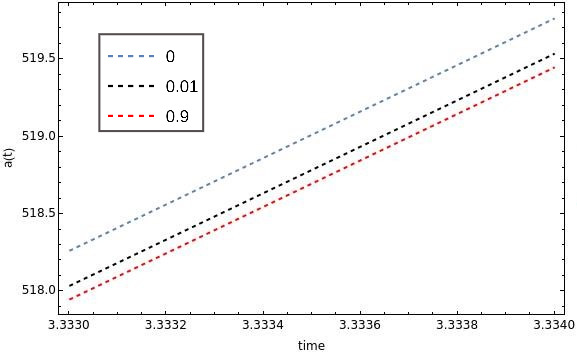}
\end{minipage}
\caption{$\textit{Left panel :}$  Plot of $a(t)$ along Y-axis vs time($t$) along X-axis for three values of $\alpha=0,0.01,0.9$ respectively in $V=e^{-k \phi}$ potential with initial condition as $a(0)=0.1$ and $\phi(0)=0.1$. 
$\textit{Right panel :} $ Magnified view.}
\label{exp.a vs t 0.1}
\end{figure}
\begin{figure}[H]
\centering
\begin{minipage}[t]{0.45\textwidth}
\centering
\includegraphics[width=1.08\textwidth]{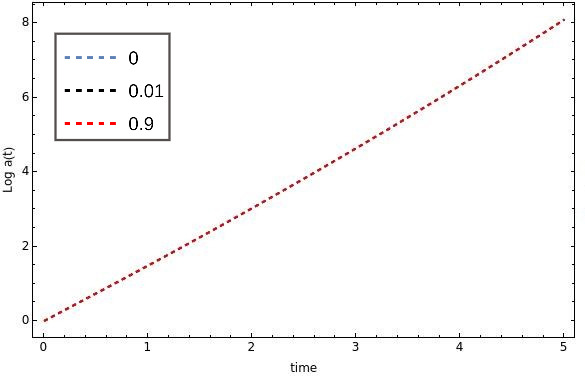}
\end{minipage}
\hfill
\begin{minipage}[b]{0.45\textwidth}
\centering
\includegraphics[width=1.15\textwidth]{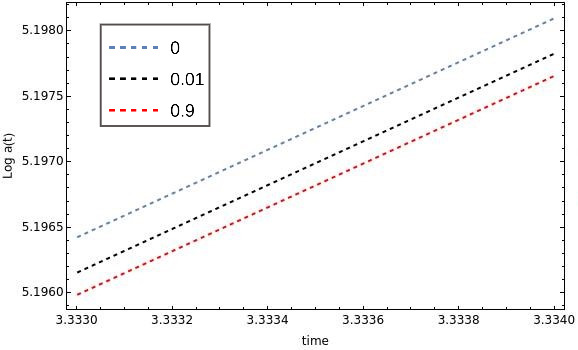}
\end{minipage}
\caption{$\textit{Left panel :}$  Plot of logarithm of $a(t)$ along Y-axis vs time($t$) along X-axis for three values of $\alpha=0,0.01,0.9$ respectively in $V=e^{-k \phi}$ potential with initial condition as $a(0)=1$ and $\phi(0)=1$.
$\textit{Right panel :} $ Magnified view.}\label{ exp. log a vs t 1}
\end{figure}
\begin{figure}[H]
\centering
\begin{minipage}[t]{0.45\textwidth}
\centering
\includegraphics[width=1.09\textwidth]{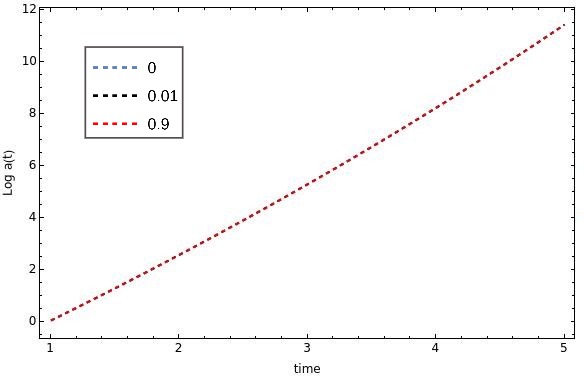}
\end{minipage}
\hfill
\begin{minipage}[b]{0.45\textwidth}
\centering
\includegraphics[width=1.15\textwidth]{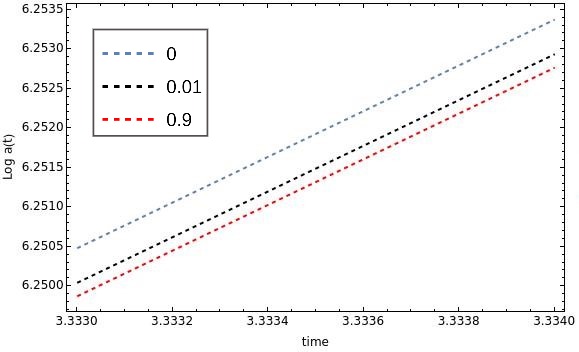}
\end{minipage}
\caption{$\textit{Left panel :}$  Plot of logarithm of $a(t)$ along Y-axis vs time($t$) along X-axis for three values of $\alpha=0,0.01,0.9$ respectively in $V=e^{-k \phi}$ potential with initial condition as $a(0)=0.1$ and $\phi(0)=0.1$.
$\textit{Right panel :} $ Magnified view. }\label{ exp. log a vs t 0.1}
\end{figure}
\begin{figure}[H]
\centering
\begin{minipage}[t]{0.45\textwidth}
\centering
\includegraphics[width=1.09\textwidth]{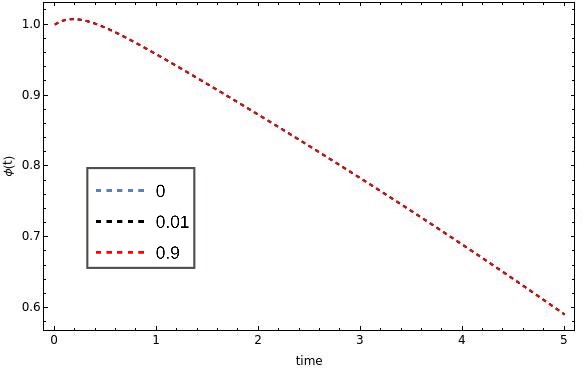}
\end{minipage}
\hfill
\begin{minipage}[b]{0.45\textwidth}
\centering
\includegraphics[width=1.15\textwidth]{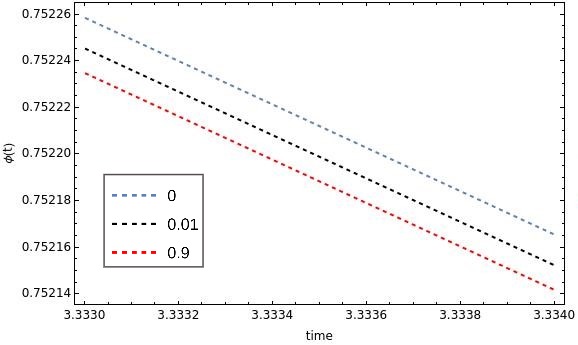}
\end{minipage}
\caption{ $\textit{Left panel :}$  Plot of $\phi(t)$ along Y-axis vs time($t$) along X-axis for three values of $\alpha=0,0.01,0.9$ respectively in $V=e^{-k \phi}$ potential  with initial condition as $a(0)=1$ and $\phi(0)=1$ . $\textit{Right panel :} $ Magnified view.}\label{phi vs t expo 1}
\end{figure}
\begin{figure}[H]
\centering
\begin{minipage}[t]{0.45\textwidth}
\centering
\includegraphics[width=1.09\textwidth]{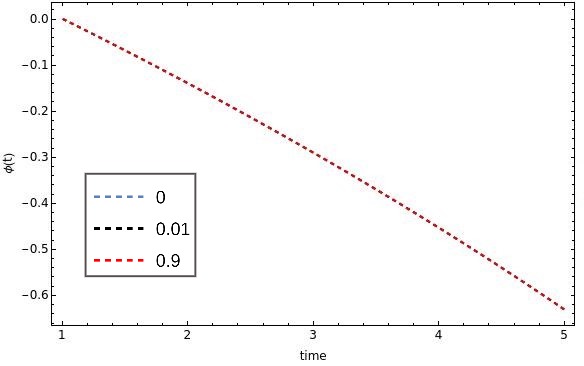}
\end{minipage}
\hfill
\begin{minipage}[b]{0.45\textwidth}
\centering
\includegraphics[width=1.15\textwidth]{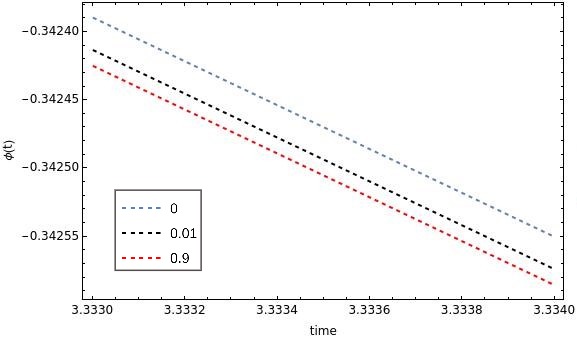}
\end{minipage}
\caption{$\textit{Left panel :}$  Plot of $\phi(t)$ along Y-axis vs time($t$) along X-axis for three values of $\alpha=0,0.01,0.9$ respectively in $V=e^{-k \phi}$ potential with initial condition as $a(0)=0.1$ and $\phi(0)=0.1$. $\textit{Right panel :} $ Magnified view.}\label{phi vs t expo 0.1}
\end{figure}
\end{widetext}

We analyze the effect of the GUP modification on the background dynamics for the exponential potential in Figs. \ref{exp.a vs t 1} - \ref{ exp. log a vs t 0.1}. The figures depict the line representing the small strength of $\alpha$, and $\alpha=0.01$ is closer to the original dynamics. However, as we increase the strength of the GUP modification, the dynamics deviate further from the original dynamics.

The behavior of the EoS parameter in the phantom field with and without the effect of GUP is represented in  Fig. \ref{ w exp.}. The EoS starts with a value of -1, indicating proximity to the cosmological constant era, and transitions toward the phantom-dominated era for both with and without GUP dynamics.

The behavior of $\phi$ for exponential potentials with different initial conditions are shown in Figs. \ref{phi vs t expo 1} and \ref{phi vs t expo 0.1}. In addition, a magnified view of the GUP deformation for different values of $\alpha$ is shown on the right-hand side of  Figs. \ref{phi vs t expo 1}) and \ref{phi vs t expo 0.1}. On the left-hand side of the scalar field, it starts from the highest value and then decreases linearly downward for different initial conditions. We observe from the graph of $\phi$ that the inclusion of a comparable level of GUP in the dynamics causes the line to distort downward from the original blue line, which represents the dynamics without GUP.
Furthermore, the plots of $|\epsilon|$ and $|\eta|$ in Figs. \ref{epsilon2} and \ref{ etaexpo} indicate inflation in the case of exponential potential.

\section{Conclusion}

In this paper, we constructed 
the GUP-corrected effective Hamiltonian from  the classical Einstein-Hilbert action. In particular,   we first considered a minimally coupled phantom scalar field with the cosmological constant as the toy model. Following this, we performed the same exercise with the presence of an arbitrary potential of the scalar field.   We focused on introducing momentum deformation to the dynamics due to GUP. Having derived the effective Hamiltonian, we obtained all the background equations of motion in terms of the Raychaudhuri, Friedmann, and Klien-Gordon equations.
Interestingly, we showed that the Klien-Gordon equation is free from any explicit quantum correction due to GUP. On the other hand, the Raychaudhuri and Friedmann equations indeed receive quantum correction explicitly. This situation is quite similar to the cosmological models constructed in the framework of LQC.


The system of equations obtained is highly nonlinear. This demands qualitative analysis using the tools of DSA to extract information about the system. We achieved this by performing a detailed DSA using the tools of linear stability analysis. We observed that the introduction of GUP affects the local behavior of the system, although the overall dynamics remain similar. This is confirmed from Fig. \ref{Q1} and Tables \ref{Table1}, \ref{Table2}, \ref{Table3} and \ref{Table4}. In the case of quadratic potential, we observed from Table \ref{Table1} and \ref{Table2} that,  after introducing GUP distortions, certain fixed points disappear. In the case of the exponential potential, Tables \ref{Table3} and \ref{Table4} indicate that, after introducing GUP corrections, the saddle point disappeared. Consequently, this leaves us with two stable points in the GUP modified scenario.

As our final goal, we returned to the question of the cosmological implications of the considered model in Sec. \ref{inflation}. We discussed inflationary scenarios after GUP corrections  as shown in Figs. \ref{a vst 1}, \ref{a vst0.1}, \ref{logavst 1} and \ref{loga vs t 0.1} for the quadratic  potential and in Figs. \ref{exp.a vs t 1}, \ref{exp.a vs t 0.1}, \ref{ exp. log a vs t 1} and \ref{ exp. log a vs t 0.1} for the exponential potential. We observed a nearly-exponential expansion for both potentials, indicating inflation occurs in these models. Finally, we calculated the slow-climb parameters for both potentials and plotted them in Fig.\ref{epsilon1}, \ref{ etasquare}, \ref{epsilon2} and \ref{ etaexpo}, which clearly show the existence of the inflationary phase. 

Furthermore, we calculated the GUP-induced EoS parameter and plotted it in Figs. \ref{wsqure} and \ref{ w exp.}, starting nearly from $-1$ and then decreasing to more negative values. To be more precise,  in the case of the quadratic potential, when we incorporate the GUP corrections, the graph of  the EoS parameter rapidly approaches the value $-1$ compared to the case without GUP corrections. We also observed a deviation from the original dynamics.

However, one serious criticism faced by the phantom field is its instability regarding perturbation analysis \cite{q5,q6}. It would be very interesting  to investigate if such instabilities are somewhat diluted due to noncommutativity corrections due to  GUP. 
In addition, it would be also very interesting  to extend our analysis for quintom fields \cite{q8,q9}, which are free from instabilities. We leave these as our future projects.

Finally we note that in this paper we restricted ourselves to the quadratic form of the GUP by setting the dimensionless parameter $\beta = 0$. However, a more general form of the GUP-modified Hamiltonian in Eq.(\ref{beta}) can be obtained by setting $\beta$ and $\epsilon$ to 1. This sets as a natural extension of our present work to be pursued in the future.



\begin{acknowledgments}
A.W. is partly supported by the US NSF  grant, PHY-2308845.
\end{acknowledgments}

\appendix
\begin{widetext}
\section{The GUP Cosmological Model}\label{ap1}

In this appendix, we apply canonical quantization to the GUP-modified Hamiltonian to derive the GUP-modified Wheeler-DeWitt (WDW) equation \cite{lopez2023generalized}. We then explore the semi-classical effects of GUP deformation, ultimately obtaining the classical Hamiltonian, which is used for dynamical system analysis.

 On substituting the deformed canonical variables Eq.(\ref{canonical}) in Eq.(\ref{HClassicalFree}) we obtain Eq.(\ref{HgupFree}) which is the GUP deformed Hamiltonian. And, using the canonical quantization, we obtain the corresponding WDW-equation as, $\hat{\mathcal{H}}\psi =0$.
Observing that the Hamiltonian is in the form of a harmonic oscillator, we can proceed with quantization using ladder operators. We consider the perturbation in the Hamiltonian to be of the order of \(\gamma^2\), so the usual ladder operators \(\tilde{a}\), \(\tilde{a}^\dagger\), and the number operator \(N\) are applicable only to the unperturbed case. For the perturbed case, new ladder and number operators are required \cite{b0}, and impose the same conditions as in the unperturbed case,
\begin{equation}
\tilde{a}|\phi_n\rangle = \sqrt{n}|\phi_{n-1}\rangle, \quad \tilde{N} |\phi_n\rangle = n|\phi_n\rangle, \quad \tilde{a}^\dagger|\phi_n\rangle = \sqrt{n + 1}|\phi_{n+1}\rangle, \quad \tilde{N} = \tilde{a}^\dagger \tilde{a}, \quad [\tilde{a}, \tilde{a}^\dagger] = 1
\end{equation}
where $|\phi_n \rangle= | \psi _n^{(0)}\rangle + \gamma|\psi_n^{(1)} \rangle+...$, and $|\psi_n^{(0)}\rangle$ is the eigenfunction of the unperturbed Hamiltonian, and $|\psi_n^{(m)}\rangle$ is the m-th order correction to the eigenstate. Since the correction $\gamma$ is very small, we can express the operators as
\begin{equation}\label{oper}
\tilde{a} = a + \sum_{n=1}^{\infty} \gamma^n \alpha_n, \quad \tilde{a}^\dagger = a^\dagger + \sum_{n=1}^{\infty} \gamma^n \alpha_n^\dagger, \quad \tilde{N} = N + \sum_{n=1}^{\infty} \gamma^n \nu_n,
\end{equation}
 we apply the above operator Eq.(\ref{oper}) to the wave function
\begin{eqnarray}
\tilde{a}|\phi_n\rangle = \left(a + \sum_{m=1}^{\infty} \gamma^m \alpha_m\right) \left(|\psi_m^{(0)}\rangle + \sum_{m=1}^{\infty} \gamma^m |\psi_n^{(m)}\rangle\right),  \quad  \tilde{a}^\dagger|\phi_n\rangle = \left(a^\dagger + \sum_{m=1}^{\infty} \gamma^m \alpha_m\right) \left(|\psi_m^{(0)}\rangle + \sum_{m=1}^{\infty} \gamma^m |\psi_n^{(m)}\rangle\right),
\end{eqnarray}
and using the procedure given in \cite{b0} we can expand the operators up to the order of $\gamma$ we get,
\begin{equation}
\tilde{a} = a - \frac{\beta \gamma}{2} \left(\frac{3 |\Lambda|}{2}\right)^{1/4} \left(a^{\dagger 3} - 6 N a^\dagger + 2 a^3\right), \quad \tilde{a}^\dagger = a^\dagger + \frac{\beta \gamma}{4} \left(\frac{3 |\Lambda|}{2}\right)^{3/4} \left(a^{\dagger 3} - 6 N a^\dagger + 2 a^3\right).
\end{equation}
Substituting these into Eq.(\ref{canonical}), after incorporating them into the expressions for momentum and coordinates, we obtain
\begin{equation}
p_{0k} = i \left(\frac{3|\Lambda|}{32}\right)^{1/4} (a_k^\dagger - a_k), \quad \tilde{q}_k = q_{0k}, \quad \tilde{k} = \left(\frac{2}{3 |\Lambda|}\right)^{1/4} (a^\dagger + a), \quad p_k = p_{0k} + \frac{i}{8} \left(\frac{3 |\Lambda|}{2}\right)^{3/4} \gamma^2 \epsilon \left(a^{\dagger 3} - 6 N a^\dagger + 2 a^3\right),
\end{equation}
the expectation values are \(\langle \tilde{q}_k \rangle = 0\) and \(\langle \tilde{p}_k \rangle = 0\). Finally, using the inverse transformation,
the volume of the universe is given by the cube of the expansion  factor as $
V = a^3(t) = \frac{3|\Lambda|}{8}(x^2 + y^2)$.
Thus, we calculate the spectrum for the volume, at the order of \(\gamma^2\), as the expected value of this expression,
\begin{align}
V(n) &= \sqrt{\frac{3}{2 |\Lambda|}} (2n + 1) + \frac{9 \beta \gamma}{8} \left(\frac{3}{8 |\Lambda|}\right)^{1/4} \left(4\sqrt{3} - \sqrt{2 |\Lambda|}\right) (1 + 2n + 2n^2) \notag \\
&\quad + \frac{3 \beta^2 \gamma^2}{64} \left(8 - 4 \sqrt{6 |\Lambda|} + 3 |\Lambda|\right) (6 + 13n + 3n^2 + 2n^3).
\end{align}

To derive the  GUP modified Wheeler-DeWitt (WDW) equation, we apply canonical quantization to the Hamiltonian described by Eq.(\ref{HgupFree}), resulting in
\begin{equation}
\hat{H}_{\text{GUP}} \psi = \hat{H}_0 \psi + i\beta\gamma \hat{H}_1 \psi + \gamma^2 \left(\frac{\beta^2}{6} + \frac{2\epsilon}{3}\right) \hat{H}_2 \psi + 2i \beta \gamma^3 \left( \frac{\beta^2 +2\epsilon}{3}\right) \hat{H}_3 \psi - \gamma^4 \left( \frac{\beta^2 +2\epsilon}{3}\right) \hat{H}_4 \psi = 0,
\end{equation}
where
\begin{align}\label{hat}
\hat{H_0} &= -\frac{1}{2}\left(\frac{\partial^2}{\partial x^2} + \frac{\partial^2}{\partial y^2}\right), \quad 
\hat{H_1} = \left(\frac{\partial^2}{\partial x^2} + \frac{\partial^2}{\partial y^2}\right)^{3/2}, \quad 
\hat{H_2} = \left(\frac{\partial^2}{\partial x^2} + \frac{\partial^2}{\partial y^2}\right)^2, \notag\\
\hat{H_3} &= \left(\frac{\partial^2}{\partial x^2} + \frac{\partial^2}{\partial y^2}\right)^{5/2}, \quad 
\hat{H_4} = \left(\frac{\partial^2}{\partial x^2} + \frac{\partial^2}{\partial y^2}\right).
\end{align}

To simplify our analysis, we will limit our considerations to the case where $\beta = 0$ and focus on terms up to the order of $\gamma^2$. To explore the semiclassical effects of the GUP deformation, we assume the wavefunction is of the form $\psi = e^{i(S(x) + G(y))}$. We then apply the operators $\hat{H}_0$ and $\hat{H}_2$ as defined in Equation (\ref{hat}), and finally, perform a WKB-approximation, we  obtain 
\begin{equation}
\hat{H}_0 \psi \approx \frac{1}{2} \left[ \left( S' \right)^2 + \left( G' \right)^2 \right] \psi + \frac{1}{2} \omega^2 \left[ x_0^2 + y_0^2 \right] \psi,
\end{equation}
\begin{equation}
\hat{H}_2 \psi \approx \left[ \left( G' \right)^4 + \left( S' \right)^4 \right] \psi + 2 \left[ \left( G' \right)^2 \left( S' \right)^2 \right] \psi,
\end{equation}
where \( S' = \frac{dS}{dx} \) and \( G' = \frac{dG}{dy} \). Now, using the usual definitions \( \frac{\partial S}{\partial x} = P_x \) and \( \frac{\partial G}{\partial y} = P_y \), we get the classical Hamiltonian
\begin{equation}
H = \frac{1}{2} \left( P_x^2 + P_y^2 \right) - \frac{3\Lambda}{8} \left( x^2 + y^2 \right) + \frac{2}{3} \gamma^2 \epsilon \left( P_x^4 + 2 P_x^2 P_y^2 + P_y^4 \right).
\end{equation}
This equation is same as Eq.(\ref{HgupFree}) for $\beta=0$ and taking only the second order of $\gamma$, which is later used to determine the GUP-deformed Friedmann, Raychaudhuri, and Klein-Gordon Equations.

\section{Proof of the Friedmann, Raychaudhuri and Klein-Gordon Equation}

In this appendix, we show the derivation of the Friedmann, Raychaudhuri and Klein-Gordon equation starting with the phantom Lagrangian. Using Eq.(\ref{FreePhantomLagrangian}) and Eq.(\ref{Hamiltonian}), and applying the Hamilton equations, we can derive the equations of motion for the system as
\begin{equation*}
P_a=\frac{\partial \mathcal{L}}{\partial \dot a}=-6a \dot a, \quad P_{\phi}=\frac{\partial \mathcal{L}}{\partial \dot \phi}=  -a^3 \dot{\phi}, \quad
\dot P_a=-\frac{\partial \mathcal{H}}{\partial a}, \quad \dot P_{\phi}=-\frac{\partial \mathcal{H}}{\partial \phi} \\
\end{equation*}
and by substituting the form of $P_a$ and $P_\phi$ so obtained from the Lagrangian equation in Eq.(\ref{Hamiltonian}).
\begin{equation*}6\dot a^2+6a\Ddot a = 3\dot a^2+\frac{3a^2\dot \phi^2}{2}+3a^2 V(\phi) + 3a^2 \Lambda,
\end{equation*}
the Raychaudhuri equation is 
\begin{equation}
 2\frac{\Ddot{a}}{a^2}+ \left(\frac{\Dot{a}}{a}\right)^2= \frac{\Dot{\phi}^2}{2} +V(\phi)+ \Lambda.
\end{equation}
The Lagrangian in Eq.(\ref{FreePhantomLagrangian}) does not depend on $\dot{N}(t)$, which implies that there is no dynamics associated with the lapse function $N(t)$. Consequently, we have $P_N \equiv \frac{\partial \mathcal{L}}{\partial \dot{N}} = 0$. Now, using Eq.(\ref{Hamiltonian}), we obtain
\begin{equation*}
3 a \dot a ^2 + \frac{\dot \phi^2 a^3}{2}- a^3( \Lambda + V)=0,
\end{equation*}
this gives the Friedmann equation as 
\begin{equation}
3H^2 =- \frac{\dot \phi^2}{2}+ \Lambda +V(\phi).
\end{equation}
Similarly, the Klein-Gordon equation is 
\begin{equation*}
3a^2 \dot a \dot \phi +a^3 \Ddot{\phi}= a^3 V'(\phi),
\end{equation*}
or,
\begin{equation}
\Ddot{\phi}+3\Dot{\phi}\frac{\Dot{a}}{a}-\frac{dV(\phi)}{d\phi}=0.
\end{equation}
\end{widetext}



\end{document}